\documentclass{elsarticle} 

\usepackage{amsmath}
\usepackage{amsthm}
\usepackage{amssymb}
\usepackage{bm}
\usepackage{xcolor}
\usepackage{graphicx}
\usepackage{caption}
\usepackage{subcaption}
\usepackage{booktabs}
\usepackage{floatrow}
\usepackage{txfonts}
\usepackage{textcomp}
\usepackage[bookmarks=false]{hyperref}
\hypersetup{colorlinks,
	linkcolor=blue,
	citecolor=blue,
	urlcolor=blue}

\usepackage{geometry}
\usepackage{numcompress}\biboptions{authoryear}

\usepackage{tikz}
\usetikzlibrary{automata, positioning, shapes.geometric, arrows}

\tikzstyle{cloud} = [ellipse,fill=white, draw=black, node distance=1cm,minimum height=2em]
\tikzstyle{box} = [rectangle, rounded corners, minimum width=4cm, minimum height=1cm,text centered, draw=black, fill=red!30]
\tikzstyle{io} = [trapezium, trapezium left angle=70, trapezium right angle=110, trapezium stretches=true, minimum width=4cm, minimum height=1cm, text centered, draw=black, fill=blue!30]
\tikzstyle{process} = [rectangle, minimum width=3cm, minimum height=1cm, text centered, draw=black, fill=orange!30]
\tikzstyle{decision} = [diamond, minimum width=3cm, minimum height=1cm, text centered, draw=black, fill=green!30]
\tikzstyle{arrow} = [thick,->,>=stealth]

\DeclareMathOperator*{\argmax}{arg\,max}
\newcommand{\Exp}[1]{\mathbb{E}\left[#1\right]}
\newcommand{\R}{\mathbb{R}}
\newcommand{\vlb}{\underline{v}}
\newcommand{\vub}{\overline{v}}
\newcommand{\vi}{v_i}
\newcommand{\vit}{v_i^\star}
\newcommand{\vj}{v_j}
\newcommand{\Fi}{F\left(\vi\right)}
\newcommand{\Wvi}{W_i\left(\vi\right)}
\newcommand{\Wvlb}{W_i\left(\vlb\right)}
\newcommand{\DWvi}{\Delta W_i\left(\vi\right)}
\newcommand{\Pvi}{P_i\left(\vi\right)}
\newcommand{\Cvi}{C_i\left(\vi\right)}

\newcommand{\Vmi}{\mathcal{V}_{-i}}
\newcommand{\Vgi}{\mathcal{V}_{>i}}
\newcommand{\Vli}{\mathcal{V}_{<i}}

\newcommand{\q}{\bm{q}_i}
\newcommand{\qgi}{\overline{q}_i}
\newcommand{\qli}{\underline{q}_i}
\newcommand{\qe}{q_{\emptyset}}
\newcommand{\Wiq}{W_q\left(\vi,\q\right)}
\newcommand{\Wfq}{W_q\left(\vi, f_q\left(\q, \xi\right)\right)}
\newcommand{\Wiqp}{W_q(\vi,\q')}
\newcommand{\Tiq}{\mathcal{T}_i^q}
\newcommand{\Siq}{\mathcal{S}_i^q}
\newcommand{\gq}{g_q(\q)}

\newcommand{\zi}{\bm{z}_i}
\newcommand{\zip}{\bm{z}'_i}
\newcommand{\zij}{z_{ij}}
\newcommand{\zijp}{z'_{ij}}
\newcommand{\se}{\sigma_{\emptyset}}
\newcommand{\slo}{\sigma_{\underline{v}_i}}
\newcommand{\shi}{\sigma_{\overline{v}_i}}
\newcommand{\Wiz}{W_z(\vi,\zi)}
\newcommand{\Wfz}{W_z\left(\vi, f_z\left(\zi, \xi\right)\right)}
\newcommand{\Wizp}{W_z(\vi,\zip)}
\newcommand{\Tiz}{\mathcal{T}_i^z}
\newcommand{\Siz}{\mathcal{S}_i^z}
\newcommand{\Liz}{\mathcal{L}_i}
\newcommand{\Lijo}{\mathcal{O}_{i,j}}
\newcommand{\Lijc}{\mathcal{C}_{i,j}}
\newcommand{\gz}{g_z(\zi)}

\newcommand{\Prob}[1]{\mathrm{Pr}\left[#1\right]}

\newcommand{\Bvi}{B_i(\vi)}
\newcommand{\Avi}{A_i(\vi)}
\newcommand{\Av}{A_i(v)}
\newcommand{\MBvi}{MB_i(\vi)}
\newcommand{\MAvi}{MA_i(\vi)}
\newcommand{\MCvi}{MC_i(\vi)}

\newtheorem{example}{Example}
\newtheorem{theorem}{Theorem}
\newtheorem{prop}{Proposition}
\newtheorem{defi}{Definition}

\begin{document}

\begin{frontmatter}

\title{Online Incentive-Compatible Mechanisms for Traffic Intersection Auctions}

\author[UNSWaddress,IFSTTARaddress]{David Rey\corref{mycorrespondingauthor}}
\cortext[mycorrespondingauthor]{Corresponding author}
\ead{d.rey@unsw.edu.au}
\author[UMNaddress]{Michael W. Levin}
\author[UNSWaddress]{Vinayak V. Dixit}

\address[UNSWaddress]{School of Civil and Environmental Engineering, UNSW Sydney, NSW, 2052, Australia}
\address[IFSTTARaddress]{Univ. Gustave Eiffel, Univ. Lyon, IFSTTAR, ENTPE, LICIT UMR\_T9401, Lyon, F-69675, France}
\address[UMNaddress]{Department of Civil, Environmental, and Geo-Engineering, University of Minnesota, Minneapolis, MN 55455, United States}

\begin{abstract}
We present novel online mechanisms for traffic intersection auctions in which users bid for priority service. We assume that users at the front of their lane are requested to declare their delay cost, i.e. value of time, and that users are serviced in decreasing order of declared delay cost. Since users are expected to arrive dynamically at traffic intersections, static pricing approaches may fail to estimate user expected waiting time accurately, and lead to non-strategyproof payments. To address this gap, we propose two Markov chain models to determine the expected waiting time of participants in the auction. Both models take into account the probability of future arrivals at the intersection. In a first model, we assume that the probability of future arrivals is uniform across lanes of the intersection. This queue-based model only tracks the number of lower- and higher-bidding users on access lanes, and the number of empty lanes. The uniformness assumption is relaxed in a second, lane-based model which accounts for lane-specific user arrival probabilities at the expense of an extended state space. We then design a mechanism to determine incentive-compatible payments in the dynamic sense. The resulting online mechanisms maximize social welfare in the long run. Numerical experiments on a four-lane traffic intersection are reported and compared to a static incentive-compatible mechanism. Our findings show that static incentive-compatible mechanisms may lead users to misreport their delay costs. In turn, the proposed online mechanisms are shown to be incentive-compatible in the dynamic sense.
\end{abstract}

\begin{keyword}
Auctions/bidding; incentive-compatibility; online mechanism design; Markov chain; traffic intersection
\end{keyword}

\end{frontmatter}

\section{Introduction}

Traffic intersections are major bottlenecks of urban transport networks. Signalized traffic intersections are typically controlled so as to improve throughput, minimize vehicle delays or reduce emissions. While these design criteria have merits they are oblivious to users' preferences. To address this limitation, auction-based mechanisms have emerged as promising alternatives to traditional traffic intersection control approaches \citep{schepperle2007agent,vasirani2012market,carlino2013auction}. In auction-based mechanisms, users are assumed to be able to declare their preferences, e.g. value of time, to an intersection manager so as to obtain services commensurate to their need. The intersection can be viewed as a server which role is to process users' service requests. In this context, the intersection manager acts as a controller which decides users' service sequence and users' payments. 

In this paper, we propose novel online auction-based traffic intersection mechanisms that allows users in the front of their lanes to bid for intersection access. We show that the proposed online mechanisms are incentive-compatible, i.e. that users cannot achieve higher utility by bidding untruthfully. Vehicles bid for individual vehicle access to the intersection, as opposed to actuating traffic signal phases which serve multiple vehicles. Such individualized control is compatible with, for example, individual-vehicle signal lighting systems (such as ramp meters) or autonomous intersection management which controls individual vehicle trajectories~\citep[e.g.][]{dresner2004multiagent, levin2017conflict}. The resulting online mechanisms are strategyproof and thus incentivize users to declare their true value of time, which is critical to promote auction-based mechanisms in urban transport networks. 

Intersection auctions have previously been explored in the context of autonomous intersection management due to the natural relationship between auction-based intersection controls and autonomous intersection management, which both decide intersection access for individual connected vehicles. We first discuss the literature on autonomous intersection management  before discussing pricing mechanisms in queueing systems and their relevance to traffic intersection auctions. We then position our paper with respect to the field and highlight our contributions.

\subsection{Autonomous intersection management}

Due to the real-time computing needs of auction-based traffic intersection mechanisms, such approaches are typically conceived in the context of autonomous intersection management (AIM), as proposed in the seminal work of \citet{dresner2004multiagent,dresner2008multiagent}. In this paradigm, traffic control is assumed to be signal-free and users are assumed to be able reserve space-time trajectories through intersections. Several works have built on and extended the AIM protocol to richer configurations. \citet{fajardo2011automated} and \citet{li2013modeling} proposed AIM protocols based on a First-Come-First-Served (FCFS) policies where vehicles are prioritized based on their arrival time at the intersection. \citet{de2015autonomous} and \citet{altche2016analysis} developed microscopic vehicle trajectory optimization formulations to coordinate vehicles through intersections. In these formulations, the intersection manager decides vehicles' service time and speed while time is discretized. \citet{zhang2016optimal,zhang2017decentralized} and \citet{malikopoulos2018decentralized} proposed decentralized approaches for traffic control based on FCFS conditions and considered energy consumption as well as vehicle separation and throughput in their formulations. \citet{mirheli2019consensus} developed a consensus-based approach for cooperative trajectory planning at signal-free intersections and shows that near-optimal solution can be achieved in competitive time. \citet{wu2019dcl} designed a multi-agent Markov decision process for cooperative trajectory planning in AIM. \citet{levin2017conflict} proposed a conflict point model that obviates the need to discretize the intersection space using tiles. This conflict point model was then adapted by \citet{rey2019blue} to accommodate both legacy and automated vehicles with signalized and signal-free traffic phases; and more recently by \citet{chen2020stability} to account for pedestrians movements. 

\subsection{Dynamic mechanism design and pricing in queueing systems}

Online auctions have been studied from the perspective of dynamic mechanism design which examines allocation mechanisms under a dynamic environment. The majority of works in dynamic mechanism design deal with sales problems which data evolve over time, such as goods with deadlines, buyers with time-varying valuations and uncertain demand \citep{bergemann2019dynamic}. \citet{kakade2013optimal} developed profit-maximizing mechanisms for environments wherein users have dynamic private valuation which build on the seminal dynamic pivot mechanism proposed by \citet{bergemann2010dynamic}. \citet{pavan2014dynamic} studied dynamic mechanisms for the case of users receiving time-varying data and decision are made over multiple time periods. \citet{mierendorff2016optimal} proposed optimal dynamic mechanisms for the case where buyers have a private deadline. The deadline of the winner determines the time of allocation hence influencing the seller's strategy. In the context of traffic intersection management, users may be assumed to have static valuations. Instead, the dynamicity of the corresponding online auctions steams from the dynamic arrival and departure of users at the intersection, which has received more attention from stochastic processes and queueing systems perspective.

Traffic intersections can be viewed as queueing systems where users wait for service, i.e. intersection access, to pursue their journey. In this regard, auction mechanisms in traffic intersections are analogous to pricing mechanisms in queueing systems, which have been extensively studied over the past few decades. \citet{naor1969regulation} was the first to study the management of queues via user payments: he considered a queueing system where arriving users faced the choice of joining a queue with a fixed payment or refrain from joining the queue. He showed that user payments could often lead to solutions maximizing social welfare. \citet{dolan1978incentive} examined a queueing system where users are serviced in a sequence determined by their declared delay costs with priority given to higher delay costs. He proposed a dynamic mechanism to determine incentive-compatible user payments. \citet{mendelson1990optimal} proposed mechanisms for optimal incentive-compatibility pricing for the M/M/1 queue. The authors considered several user classes that differ by their delay costs and identify optimality conditions for net value maximization under elastic demand. In a similar context, \citet{bradford1996pricing} studied incentive-compatibility in the context of multiserver queues. \citet{kittsteiner2005priority} studied a priority queueing system where users have varying service (processing) times and where queue lengths are unobservable. Users are allowed to bid for service and are privately informed of their processing time. In this incomplete information context, they show that the shortest processing time (SPT) discipline is approximately efficient if the curvature of the cost function is relatively low. Further details on equilibrium behavior in queueing systems can be found in a review by \citet{hassin2003queue} and more recent works are surveyed by \citet{hassin2016rational}.
	
\subsection{Traffic intersection auctions}

Traffic intersection auctions have received a growing attention over the past few years, notably with the advent of connected and automated vehicle technology. \cite{schepperle2007agent} proposed a subsidy-based mechanism for slot allocation which aim to balance vehicle waiting time. First-In First-Out (FIFO) constraints are accounted for but incentive-compatibility is not guaranteed. The ``effect of starvation'' and its impact on fairness are discussed. \cite{vasirani2012market} developed a policy based on combinatorial auctions for the allocation of reservations at traffic intersections. The auction winner pays a price that is exactly the bid that was submitted, which is not incentive-compatible. \cite{carlino2013auction} proposed several types of auctions where users can bid on phases or reservations (one vehicle at a time). All drivers can participate in the auction but the candidates only include drivers at the front of their lane. Winners split the cost of the second-highest bid with proportional payment, which yields a static incentive-compatible mechanism and the strategyproofness of the dynamic case is not guaranteed. \cite{sayin2018information} proposed an auction protocol and mechanism in which all vehicles within range of the intersection manager communicate their value of time, and the intersection manager assigns reservations to those vehicles via a static auction. The proposed mechanism is incentive-compatible in the static sense, i.e. assuming all participants in the auction are known at the time payments are determined. \cite{censi2019today} introduced a credit-based auction mechanism, i.e. a karma system in which agents pay other agents karma for priority. Agents with a low priority today have an incentive to lose bids so they can achieve higher priority tomorrow by acquiring more karma. The authors attempt to calculate a Nash equilibria of user bids, and they show that a centralized strategy can be more unfair than some of the Nash equilibria identified. Recently, \cite{lin2020pay} proposed a mechanism for pricing intersection priority based on transferable utility games. In each game, players have the possibility to trade time among themselves and ``winners pay losers to gain priority''. The authors provide empirical evidence that their approach is robust against adversarial user behavior, but no formal proof of incentive-compatibility is presented. 

A synthesis of the state-of-the-art on traffic intersection auctions is presented in Table \ref{tab:lit}. The column \textbf{Mechanism} describes the mechanism of the corresponding paper. The column \textbf{Auction} indicates the type of auction: static or dynamic. A static auction is an auction in which the set of participants is assumed known, whereas a dynamic auction allows participants to enter and leave the auction dynamically. Observe that, in a static traffic intersection auction, losers may need participate in multiple auction rounds until being serviced during which new users may arrive and further delay these losers. The column \textbf{IC} indicates whether the mechanism is incentive-compatible or not. The column \textbf{F} indicates if fairness is considered in the mechanism, either to some extent (some) or via constraints within the mechanism (yes). The column \textbf{E} indicates if the mechanism is efficient, i.e. if social welfare is optimized. The column \textbf{C} indicates if intersection capacity is considered in the mechanism, either to some extent (some) or explicitly optimized (yes). Table \ref{tab:lit} highlights that, to the best of our knowledge, existing mechanisms for traffic intersection auctions are restricted to static auctions. This synthesis also emphasizes that fairness considerations and intersection capacity are seldom optimized jointly, thus underlining existing research gaps in the design of traffic intersection auction mechanisms.

\begin{table}
\centering
\begin{tabular}{lllllll}
\toprule
\textbf{Paper} & \textbf{Mechanism} & \textbf{Auction} & \textbf{IC} & \textbf{F} & \textbf{E} & \textbf{C} \\
\midrule
\cite{schepperle2007agent} & subsidy-based & static & no & some & no & some \\
\cite{vasirani2012market} & combinatorial auction & static & no & no & yes & yes \\
\cite{carlino2013auction} & marginal cost & static & yes & yes & yes & some \\
\cite{sayin2018information} & marginal cost & static & yes & no & yes & yes \\
\cite{censi2019today} & karma system & static & no & yes & yes & some \\
\cite{lin2020pay} & transferable utility & static & no & some & yes & yes \\
This paper & expected marginal cost & dynamic & yes & no & yes & some \\
\bottomrule
\end{tabular}
\caption{Summary of the state-of-the-art on traffic intersection auctions. The column \textbf{Mechanism} describes the mechanism; \textbf{Auction} indicates the type of auction; \textbf{IC} indicates whether the mechanism is incentive-compatible; \textbf{F} indicates if fairness is considered in the mechanism; \textbf{E} indicates if the mechanism is efficient, i.e. social welfare is optimized; \textbf{C} indicates if intersection capacity is considered or optimized.}
\label{tab:lit}
\end{table}

\subsection{Our contributions}

As highlighted in our review of the literature, existing payment mechanisms for traffic intersection auctions are either only incentive-compatible for static auctions, or dynamic but not incentive-compatible. Static auctions do not account for future arrivals when determining user payments. As we will show in our numerical experiments, static payment mechanisms may incentivize users to misreport their preferences by bidding lower than their true delay costs and obtaining a higher utility, thus potentially compromising the acceptability of the corresponding mechanisms. Since users are expected to arrive dynamically at traffic intersections, online incentive-compatible mechanisms are required to ensure truthful user behavior in the long run.

We make the following contributions to the field. We propose two online payment mechanisms for traffic intersection auctions. The two proposed mechanisms use Markov Chains (MC) to model the expected waiting time of users in the system but differ in the way the expected waiting time of users is determined. First, a queue-based model which assumes uniform arrival rates across the intersection is proposed and shown to be computationally inexpensive. This approach is then extended to a lane-based model which is able to capture non-uniform arrival rates across the lanes of the intersection. We show that both online mechanisms are incentive-compatible in the dynamic sense. We conduct numerical experiments to explore the behavior of the proposed online mechanisms and quantify the benefits of both queue-based and lane-based mechanisms. Finally, our simulations highlight the limitations of static incentive-compatible payment mechanisms compared to the proposed dynamic approaches. 

We introduce the traffic intersection auction in Section \ref{auction} and present the proposed online mechanisms in Section \ref{mechanisms}. Numerical experiments are reported in Section \ref{num} and conclusions are summarized in Section \ref{con}.

\section{Dynamic traffic intersection auction framework}
\label{auction}

We consider a traffic intersection with a finite number of access lanes. We consider discrete time and we assume that at each time period, users arrive on access lanes with a known probability. Upon arriving at the front of their lane, users are requested to declare their delay cost which will be used as their bid in a combinatorial auction. We assume that users seek to minimize a generalized cost function which is a linear combination of their expected waiting time and their payment to the intersection manager. We assume that users are serviced sequentially by a single server, i.e. the traffic intersection and that service is nonpreemptive, i.e. cannot be interrupted once started. This does not prevent several users to simultaneously traverse the intersection. Instead, the proposed traffic intersection auction only aims to determine the sequence in which users are to be serviced and the payments they should be charged based on their declared preferences.\newline

The goal of the intersection manager is to maximize social welfare which is defined as the total generalized user cost. To achieve optimal social welfare, we seek to determine incentive-compatible user payments to ensure the truthful declaration of their delay costs. To ensure truthful user behavior, traffic intersection lanes are serviced in order of decreasing declared user delay costs. The expected waiting time of users is a function of the declared delay costs and the state of the system. In turn, the payment of users is to be determined by the intersection manager so as to maximize social welfare and ensure truthfulness. 

To present the proposed online mechanism for determining incentive-compatible payments, we consider the case of a user $i$ arriving at the front of its lane and declaring a delay cost of $\vi$. Let $Q$ be the number of access lanes of the intersection. We denote $\qgi$ (resp. $\qli$) the number of lanes which are occupied by users with greater (resp. lower) delay cost than $i$. Further, we denote $\qe$ the number of empty lanes. Summing $\qgi$, $\qli$ and $\qe$ corresponds to the $Q-1$ lanes of the intersection which are not occupied by user $i$. Hence, the following equation holds for any user $i$ at any time period:
\begin{equation}
Q = \qli + \qgi + \qe + 1.
\label{eq:Q}
\end{equation}

Accordingly, we can define the pricing queue of the proposed dynamic traffic intersection auction.

\begin{defi}[Pricing queue]
In a traffic intersection with $Q$ access lanes, the \textit{pricing queue} is a dynamic set of at most $Q$ users which represents the users at the front of their lane at a given time period. 
\end{defi}

The pricing queue consists of the set of users which have no users in front of them in their access lane to the intersection. In the proposed traffic intersection auction, only users in the pricing queue are participants. This implies that users who are behind other users are not eligible to participate until they reach the front of their lane. This auction design aims to obviates blocking effects induced by non-overtaking conditions in access lanes of the intersection. Under such typical, non-overtaking conditions users waiting behind other users in their lane cannot bypass users in front of them, thus preventing them to be serviced before reaching the front of their lane. In the proposed approach, only users in the pricing queue participate in the auction. While this dynamic auction framework restricts the number of participants to at most $Q$ users, every user queueing in the intersection will eventually reach the front of its lane and participate in a single instance of the auction. This is in contrast to approaches in the literature where auction ``losers'' may participate in several auctions. We refer readers to \citet{carlino2013auction} who presented pricing mechanisms for traffic intersection auctions in which blocked users are allowed to ``vote'' for their lane. \newline

The objective of the proposed traffic intersection auction is to determine incentive-compatible user payments upon users joining the pricing queue. Let $\Wvi$ be the expected waiting time of user $i$ declaring a delay cost of $\vi$. Let $\vit$ be the true delay cost of user $i$, the expected waiting time cost of user $i$ declaring a delay cost of $\vi$ is $\vit \Wvi$. Let $\Pvi$ be the payment of user $i$ declaring $\vi$, which is to be determined by the intersection manager. We denote $\Cvi$ the generalized cost for user $i$ and define the user objective function as:

\begin{defi}[User objective function]
The objective function of user $i$ is:
\begin{equation}
\min_{\vi} \Cvi = \min_{\vi} (\vit \Wvi + \Pvi).
\label{eq:obj}
\end{equation}
\end{defi}

Following the approach of \cite{dolan1978incentive}, we will show that setting the payment $\Pvi$ equal to the expected marginal delay cost user $i$ imposes to other users is incentive-compatible, i.e. the user objective is minimal for $\vi = \vit$. \newline

We assume that the intersection manager has knowledge of the probability distribution of users' valuation (delay costs), and of the arrival probability of users on access lanes of the intersection. These assumptions are reasonable since one can assume that the intersection manager can observe user behavior (arrival rate and valuations) and learn these probabilities over time. Let $\Fi$ be the cumulative distribution function representing the probability that a user declares a delay cost $v \leq \vi$. Further, let $\vlb$ and $\vub$ be lower and upper bounds on users' delay costs. For conciseness, we hereby use the term \textit{lower-bidding user} (resp. \textit{higher-bidding user}) to refer to a user with a declared delay cost lower (resp. higher) than that of user $i$. 

An overview of the proposed dynamic auction process is summarized in Figure 1. Users are assumed to arrive dynamically and submit their bids (delay cost) to the intersection manager. The intersection manager then determines a user payment $\Pvi$ based on the user's bid $\vi$ using either the proposed queue- or lane-based model.  The user is then charged $\Pvi$ and moves through the intersection.

\begin{figure}
	\centering
	\begin{tikzpicture}[node distance=1.5cm]
		\node (start) [cloud]{Start};
		\node (idle) [box, text width=3cm, below of=start] {Intersection manager server idle};
		\node (new) [decision,below of=idle,yshift=-0.5cm] {New user?};
		\node (in1) [io, below of=new,yshift=-0.5cm] {User $i$ bids $\vi$};
		\node (pro1) [process, text width=4cm, below of=in1] {Determine payment $\Pvi$ using Queue- or Lane-based model};
		\node (ou1) [io, text width=3cm, below of=pro1, yshift=-0.2cm] {User $i$ is charged $\Pvi$ and moves through intersection};
		\draw [arrow] (start) -- (idle);
		\draw [arrow] (idle) -- (new);
		\draw [arrow] (new) -- ++ (2.5cm,0cm) node[near start,anchor=south] {No} |- (idle);
		\draw [arrow] (new) -- node[anchor=east] {Yes} (in1);
		\draw [arrow] (in1) -- (pro1);
		\draw [arrow] (pro1) -- (ou1);
		\draw [arrow] (ou1) -- ++ (4cm,0cm) |- (idle);
	\end{tikzpicture}
	\caption{Online traffic intersection auction process}
	\label{fig:flow}
\end{figure}
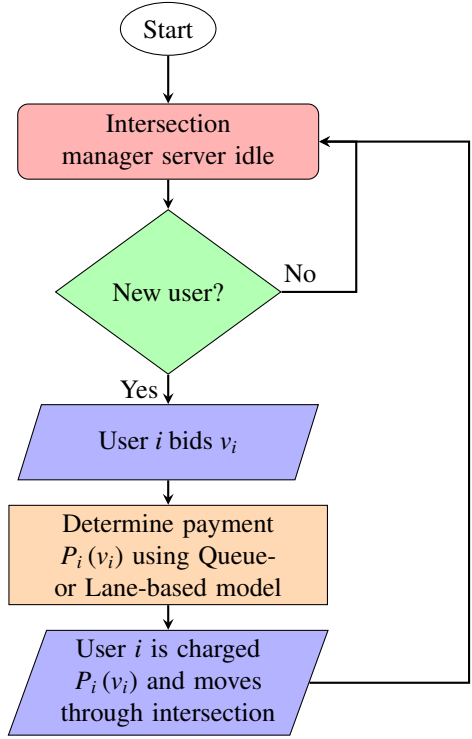

We next introduce the proposed online mechanisms to determine dynamic incentive-compatible payments for traffic intersection auctions.

\section{Online incentive-compatible mechanisms}
\label{mechanisms}

We propose two online mechanisms to determine the expected waiting time and payment of users. The proposed mechanisms differ in the way the expected waiting time of users is determined. We first consider a queue-based model which only tracks the number of users bidding higher or lower than the reference user (Section \ref{queue}). We next extend this model to a lane-based approach which tracks users' arrival lane (Section \ref{lane}). A generic payment mechanism is proposed to determine incentive-compatible payments for both queue- and lane-based models which result in two online mechanisms (Section \ref{payment}). The queue-based mechanism assumes that the arrival probability of users is uniform across lanes of the intersection and provides a relatively low-dimensional mathematical framework. This uniformness assumption is relaxed in the lane-based mechanism which can accommodate non-uniform lane arrival probabilities at the expense of an extended state space. 

\subsection{Queue-based model}
\label{queue}

We use a MC to determine the expected waiting time of user $i$ under different system states. The state of the queuing system w.r.t. user $i$ is represented by the number of lanes occupied by lower bidding vehicles lanes $\qli$ the number of empty lanes $\qe$. We denote $\q$ the vector $\q = (\qli,\qe)$ characterizing the state of the system w.r.t. user $i$. User $i$ is serviced when all lanes are either occupied by a lower-bidding user or empty. Observe that once a lane is occupied by a lower-bidding user w.r.t. user $i$, it remains in this state until $i$ is serviced. Hence, from the perspective of user $i$, the state space of the MC can be characterized by the inequality $\qli+\qe\leq Q-1$. This leads to the following characterization of the state space.

\begin{defi}[State space in the queue-based model]
In the queue-based model, the state space of the MC for user $i$ declaring a delay cost of $\vi$ in a pricing queue of size $Q$ is:
\begin{equation}
\Siq=\left\{\q=(\qli,\qe) \in \mathbb{N}^2: \qli + \qe \leq Q-1 \right\}.
\end{equation}  
\end{defi}

For each state $\q \in \Siq$, we are interested in determining the cost-to-go from this state through the MC. If $\qli + \qe = Q-1$, then \eqref{eq:Q} yields $\qgi = 0$ which is equivalent to require that the pricing queue does not contain any user with declared delay costs higher than that of $i$. Hence, if $\qli + \qe = Q-1$, then the MC converges and user $i$ is serviced. Hence, to determine the state transition probabilities from state $\q$ to state $\q'$, we assume that $\qli+\qe<Q-1$.

To calculate the expected waiting time of user $i$ in the queue-based model, we enumerate system states with varying number of lower-bidding and empty lanes w.r.t. user $i$. We wish to define a function $f_q$ which gives the transition between the state at time step $t$ to the state at time $t+1$. The transition function of the MC, $f_q$, depends on the realization of random variables. For this problem, there are two random variables affecting the transition: user delay costs, a continuous random variable; and whether a new vehicle arrives on each lane, a boolean value. Hence, the uncertain data of the problem can be represented by a vector $\xi \in \R^Q \times 2^Q$ where $R^Q$ represents user delays costs and $2^Q$ represents vehicle arrivals. Since the state space of the queue-based model is $\Siq$, the transition function maps the current state $\q \in \Siq$ and the uncertain data $\xi \in \R^Q \times 2^Q$ to the next state, i.e. $f_q :\Siq\times \R^Q \times 2^Q \rightarrow \Siq$. We denote $\gq$ the one-step-cost of the MC representing the service time of the system in state $\q$ and we assume that the one-step-cost is deterministic. This assumption is plausible since users' behavior is assumed to be deterministic, i.e. users cannot change their declared delay cost over time. 

Let $\Wiq$ be the expected waiting time of user $i$ declaring $\vi$ for state $\q$, which represents the cost-to-go of the MC representing the state evolution in the queue-based model. Let $\Tiq = \{\q \in \Siq : \qe + \qli = Q - 1\}$ be the set of terminal states in the queue-based mechanism. The expected waiting time of user $i$ declaring $\vi$ for state $q$ can be determined as:
\begin{equation}
\Wiq =\begin{cases}
0 & \text{if } \q \in \Tiq, \\
\Exp{\Wfq} + \gq  & \text{otherwise}.
\end{cases} 
\label{eq:wq}
\end{equation}

Note that the term $\Exp{\Wfq}$ represents the expected value of the expected waiting time $\Wfq$ after the transition function $f_q$. If $\q$ is not a terminal state, i.e. $\qli+\qe < Q-1$, the expected waiting time of user $i$ can be calculated using the probability $\Prob{\q'|\q}$ of transitioning from state $\q$ to state $\q'$. The expected waiting time of user $i$ declaring $\vi$ is:
\begin{equation}
\Exp{\Wfq} = \sum\limits_{\q' \in \Siq} \Prob{\q'|\q} \Wiqp.
\label{eq:ewt}
\end{equation}

Combining Eqs. \eqref{eq:wq} and \eqref{eq:ewt} yields the following system of linear equations:
\begin{subequations}
	\begin{align}
	\Wiq &= 0, \quad &\forall \q \in \Tiq, \\
	\Wiq (1 - \Prob{\q|\q}) - \sum_{\q' \in \Siq : \q' \neq \q} \Wiqp \Prob{\q'|\q} &= \gq, \quad &\forall \q \in \Siq \setminus \Tiq.
	\end{align}
	\label{eq:wq2}
\end{subequations}

Let $p$ be the probability that a user arrives on a lane of the intersection at the next time period. Recall that $\Fi$ is the cumulative distribution function of users bids. On each lane, three events may occur with the following probabilities: i) a lane can become or remain empty with probability $(1-p)$; ii) a lower-bidding user can arrive with probability $p\Fi$; and iii) a higher-bidding user can arrive with probability $p(1-\Fi)$.	Observe that there are $\binom{\qe+1}{\qe'}$ ways to choose $\qe'$ empty lanes among $\qe+1$ lanes. There are $\qli' - \qli$ lanes which become occupied by lower-bidding users and there are $\binom{\qe+1-\qe'}{\qli'-\qli}$ ways to choose $\qli'-\qli$ of the remaining $\qe+1-\qe'$ non-empty lanes to become occupied by a lower-bidding user. Accordingly, the transition probability from state $(\qli,\qe)$ to state $(\qli',\qe')$ is:
\begin{equation}
\Prob{\q'|\q}= \binom{\qe+1}{\qe'}(1-p)^{\qe'} \binom{\qe+1-\qe'}{\qli'-\qli} (p\Fi)^{\qli'-\qli} (p(1-\Fi))^{\qe+1-\qe'- \qli'+\qli}.
\label{eq:prob}
\end{equation}

The expected waiting time defined in Eq. \eqref{eq:wq} is a recursive function which can be calculated by solving a series of systems of linear equations where each equation corresponds to a possible state $\q\in \Siq$. Since all states such that $\qli+\qe=Q-1$ are terminal, their corresponding expected waiting time is null and these variables can be eliminated from the system of equations. Hence, we need only to consider the set of pairs of nonnegative integers $(\qli,\qe)$ such that $\qli + \qe \leq Q-2$. Observe that each variable $\qli$ or $\qe$ can take $Q-2+1$ values, thus the number of equations in the system is $\sum_{k=0}^{Q-1} k = \binom{Q}{2}$.

Further, observe that $\qli$ can never decrease while $\qe$ can increase or decrease. Hence, for any state $\q=(\qli,\qe)$, we can calculate $\Wiq$ in a dynamic programming fashion starting with the maximum number of lanes, i.e. $k=Q-1$, and iterating downwards until $k=\qli$. At each iteration $k$, we solve the system of $Q-k$ linear equations with $Q-k$ variables corresponding to states $\q=(k,l)$ for $0\leq l \leq Q-1-k$. 

We illustrate the queue-based model in Example \ref{ex:queue}.

\begin{example}
\label{ex:queue}
Consider an intersection with $Q=3$ access lanes. The space state of the MC can be represented as in Figure \ref{fig:mc_queue} where the bottom row corresponds to terminal states.

\begin{figure}[!b]
\centering
\begin{tikzpicture}
\tikzset{node style/.style={state, 
		minimum width=1cm,
		line width=0.3mm,
		fill=gray!20!white}}
\node[node style] at (7, 0) (00) {$(0,0)$};
\node[node style] at (5, -2) (10) {$(1,0)$};
\node[node style] at (9, -2) (01) {$(0,1)$};
\node[node style] at (3, -4) (20) {$(2,0)$};
\node[node style] at (7, -4) (11) {$(1,1)$};
\node[node style] at (11, -4) (02) {$(0,2)$};
\draw[every loop,auto=right,line width=0.3mm]
(00)     edge[loop above]            node {} (00)
(10)     edge[loop above]            node {} (10)
(01)     edge[loop above]            node {} (01)
(00)     edge[bend right=10]            node {} (10)
(00)     edge[bend right=10]            node {} (01)
(10)     edge[bend right=10]            node {} (20)
(10)     edge[bend right=10]            node {} (11)
(01)     edge[bend right=10]            node {} (00)
(01)     edge[bend right=10]            node {} (10)
(01)     edge[bend right=10]            node {} (20)
(01)     edge[bend right=10]            node {} (11)
(01)     edge[bend right=10]            node {} (02);
\end{tikzpicture}
\caption{Illustration of the MC diagram for the queue-based model corresponding to 3-lane intersection ($Q=3$). The number of states is $|\Siq|=6$. The states are written in the form $(\qli,\qe)$.}
\label{fig:mc_queue}
\end{figure}
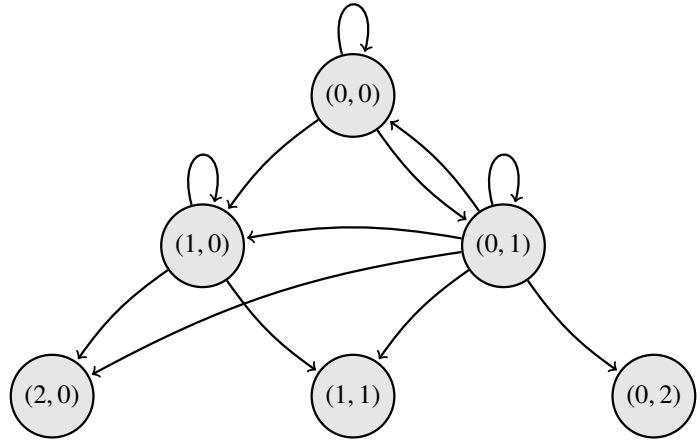

The reachable sets denoted $\Siq\left(\qli,\qe\right)$ of the three non-terminal states of this MC are:
\begin{align*}
\Siq(0,0) &= \{(0,0),(0,1),(1,0)\}, \\
\Siq(0,1) &= \{(0,0),(0,1),(0,2),(1,0),(1,1),(2,0)\}, \\
\Siq(1,0) &= \{(1,0),(1,1),(2,0)\}. 
\end{align*}

We now illustrate the behavior of the queue-based model when determining the expected waiting time for the state $\q=(1,0)$.

\begin{enumerate}
\item At the first iteration, $k=Q-1=2$ and there are $Q-k=1$ equation corresponding to variable $W_q(\vi,(2,0))$, which is trivial since $(2,0)$ is a terminal state, i.e. $W_q(\vi,(2,0)) = 0$. 

\item At the second iteration, $k=Q-2=1$ and there are $Q-k=2$ equations corresponding to variables $W_q(\vi,(1,0))$ and $W_q(\vi,(1,1))$, i.e.:
\begin{align*}
W_q(\vi,(1,0)) \left(1 - \Prob{(1,0)|(1,0)}\right) - W_q(\vi,(1,1)), \Prob{(1,1)|(1,0)}&\\
 - W_q(\vi,(2,0)) \Prob{(2,0)|(1,0)} &= g_q((1,0)), \\
W_q(\vi,(1,1))  &= 0.
\end{align*}

Since the states $(2,0)$ and $(1,1)$ are terminal states, the above system collapses to:
\[
W_q(\vi,(1,0)) = \frac{g_q((1,0))}{\left(1 - \Prob{(1,0)|(1,0)}\right)}.
\]
\end{enumerate}

Assuming a uniform distribution of users' delay costs in the range $\vlb = $\$5 /hour and $\vub = $\$10 /hour, i.e. $\mathcal{U}(5,10)$, and a one-step-cost of 1 s, i.e. $g_q((\qli,\qe)) = 1$. If the arrival probability is $p=1/3$ and if user $i$ has a value of time of 7 \$/hour corresponding to a bid of $\vi = 0.19$~\textcent, the expected waiting time of state $\q = (0,1)$ is $W_q(\vi,(1,0)) = 1.25$ s.
\end{example}

\subsection{Lane-based model}
\label{lane}

We now consider an alternative approach where lane arrival probabilities can be non-uniform across lanes of the intersection. We abuse notation and denote $p_j$ the probability of a user arriving on lane $j$ at the next time period. Let $\Liz = \{\se,\slo,\shi\}$ be the set of possible lane-based states with regards to user $i$ bidding $\vi$; where $\se$ represents an empty lane, $\slo$ the arrival of a user bidding lower than $\vi$ and $\shi$ the arrival of a user bidding higher than $\vi$. Let $\zij \in \Liz$ be the state of lane $j$ w.r.t. user $i$. We denote $\zi = (z_{i,1}, \ldots, z_{i,Q-1})$ the state w.r.t. user $i$ in the lane-based model. 

\begin{defi}[State space for lane-based model]
In the lane-based model, the state space of the MC for user $i$ declaring a delay cost of $\vi$ in a pricing queue of size $Q$ is
\begin{equation}
\Siz = \left\{\zi = (z_{i,j_1},\ldots,z_{i,j_{Q-1}}) \in \Liz^{Q-1}\right\}.
\end{equation}
\end{defi}

In the lane-based model, the number of possible states w.r.t. user $i$ in an intersection with $Q$ lanes is $|\Liz|^{Q-1}$, hence a four-lane intersection has $3^3 = 27$ possible states. 

Let $f_z :\Siz\times \R^Q \times 2^Q\rightarrow \Siz$ be the MC transition function of the lane-based model and let $\gz$ be the one-step cost. Let $\Tiz \subset \Siz$ be the set of terminal states, i.e. $\Tiz = \{\zi \in \Siz: \zij = \se \vee \zij = \slo, j=1,\ldots,Q-1\}$. The expected waiting time $\Wiz$ of user $i$ declaring $\vi$ for state $\zi$ in the lane-based model is thus:
\begin{equation}
\Wiz = \begin{cases}
0 & \text{if } \zi \in \Tiz,\\
\Exp{\Wfz} + \gz  & \text{otherwise},
\end{cases} 
\label{eq:wzi}
\end{equation}

with:
\begin{equation}
\Exp{\Wfz} = \sum_{\zip \in \Siz} \Prob{\zip|\zi} \Wizp.
\label{eq:wfzi}
\end{equation}

This leads to the following system of linear equations:
\begin{subequations}
\begin{align}
\Wiz &= 0 \quad &\forall \zi \in \Tiz, \\
\Wiz (1 - \Prob{\zi|\zi}) - \sum_{\zip \in \Siz : \zip \neq \zi} \Wizp \Prob{\zip|\zi} &= \gz \quad &\forall \zi \in \Siz \setminus \Tiz.
\end{align}
\label{eq:WTZ}
\end{subequations}

Unlike the systems of equations \eqref{eq:wq2}, the system \eqref{eq:WTZ} does not admit an intuitive decomposition which could be used to solve the system of equations via a recursive algorithm. Thus, in our numerical experiments, \eqref{eq:WTZ} is solved in a single step using traditional linear algebra codes. 

To determine transition probabilities in the lane-based mechanism, it is necessary to track which user will be serviced at the next time period. For any non-terminal state, there is at least one higher-bidder in the pricing queue, i.e. $\qgi \geq 1$, and the highest-bidder in the queue will be serviced next. Let $\Vgi$ be the set of higher-bidders, i.e. $\Vgi = \{j_1, j_2, \ldots, j_{\qgi}\}$ with $\vi < v_{j_1} < v_{j_2} < \ldots < v_{j_{\qgi}}$. If there is more than a single user bidding higher than user $i$, i.e. $\qgi > 1$, it is unknown which of these higher-bidders will be the highest-bidder. Since users' delay costs are assumed to all follow the same probability distribution, the probability that a higher-bidder is the highest-bidder is uniform, i.e. $1/\qgi$.

Let $\zij(k)$ be the state of lane $j$ with regards to user $i$ assuming lane $k$ is the moving lane, i.e. the lane occupied with the highest-bidder in the pricing queue. Let $\Prob{\zijp|\zij(k)}$ be the transition probability of lane $j$ with regards to user $i$ from state $\zij(k)$ to state $\zijp$. Since the arrival process of a lane is assumed to be independent of that of other lanes, the transition probability from state $\zi$ to $\zip$ denoted $\Prob{\zip|\zi}$ is
\begin{equation}
\Prob{\zip|\zi} = \Prob{(z'_{i,1}, \ldots, z'_{i,Q-1})|(z_{i,1}, \ldots, z_{i,Q-1})} = \frac{1}{\qgi} \sum_{k \in \Vgi}  \prod_{j=1}^{Q-1} \Prob{\zijp|\zij(k)}.
\label{eq:dprob}
\end{equation}	

In the lane-based mechanism, transition probabilities are function of current lane state. If lane $j$ is empty or if lane $j$ is occupied by the highest bidder in the pricing queue, then we say that this lane is open. Otherwise, lane $j$ is either occupied by a lower-bidding user or a higher-bidding user who will not be serviced next and we say that his lane is closed. Let $\Vmi$ be the set of users in the pricing queue other than user $i$. Let $\Lijo \subseteq \Liz$ and $\Lijc \subseteq \Liz$ be the set of open and closed states for lane $j$ with regards to user $i$, respectively:
\begin{align}
\Lijo = \begin{cases}
\{\se,\shi\} &\text{ if } j \in \argmax\{v_k : k \in \Vmi\}, \\
\{\se\} &\text { otherwise},
\end{cases}\\
\Lijc = \begin{cases}
\{\slo\} &\text{ if } j \in \argmax\{v_k : k \in \Vmi\}, \\
\{\slo,\shi\} &\text { otherwise}.
\end{cases}
\end{align}

For any lane $j$ in state $\sigma \in \Lijo$, we have the following transition probabilities:
\begin{subequations}
\begin{align}
\Prob{\se|\sigma} &= 1-p_j \quad &&\forall j \in \{1,\ldots,Q-1\}, \forall \sigma \in \Lijo,\\
\Prob{\slo|\sigma} &= p_j\Fi \quad &&\forall j \in \{1,\ldots,Q-1\}, \forall \sigma \in \Lijo,\\
\Prob{\shi|\sigma} &= p_j(1 - \Fi) \quad &&\forall j \in \{1,\ldots,Q-1\}, \forall \sigma \in \Lijo.
\end{align}
\end{subequations}

Otherwise if lane $j$ is in state $\sigma \in \Lijc$, this lane remains in its current state with probability one, since the corresponding user cannot be serviced at the next time period. 
\begin{align}
\Prob{\sigma'|\sigma} &= 
\begin{cases}
1 \quad \text{ if } \sigma' = \sigma,\\
0 \quad \text{ otherwise},
\end{cases}
\quad &&\forall j \in \{1,\ldots,Q-1\}, \forall \sigma \in \Lijc.
\end{align}

Using these lane-based transition probabilities, we can determine intersection-based transition probabilities via Eq. \eqref{eq:dprob}.\newline

The queue-based model can be viewed as a special case of the lane-based model. Let $\bm{1}_{\zij=\sigma}$ be the indicator function taking value 1 if $\zij=\sigma$ and 0 otherwise. We first define the mapping between the state spaces of both queue- and lane-based models.

\begin{defi}
Let $h_i : \Siz \rightarrow \Siq$ be a function mapping the state space of the lane-based model to the state space of the queue-based model: $h_i : \zi \mapsto \q = h_i(\zi)$ with $\qli = \sum_{j=1}^{Q-1} \bm{1}_{\zij = \slo}$ and $\qe = \sum_{j=1}^{Q-1} \bm{1}_{\zij = \se}$.
\end{defi}

The proposition below highlights the correspondence between queue- and lane-based models.

\begin{prop}
If lane arrival probabilities are uniform, i.e. $p = p_j$ for all $j=1,\ldots Q-1$, and if the one-step-costs of the queue- and lane-based models are such that $g_z(\zi) = g_q(\q)$ for all states $\zi \in \Siz$ and $\q \in \Siq$ such that $\q = h_i(\zi)$, then $\Wiz = W_q(\vi,\q)$ for all such states.
\begin{proof}
The probability of state $\q$ occurring is:
\begin{equation}
\Prob{\q} = (1-p)^{\qe} p^{\qli+\qgi} \Fi^{\qli} (1-\Fi)^{\qgi}.
\end{equation}

The probabilities of occurrence for lane-specific states $\zij$ are:
\begin{equation}
\Prob{\zij} = \begin{cases}
1 - p_j &\text{ if } \zij = \se, \\
p_j \Fi &\text{ if } \zij = \slo, \\
p_j (1-\Fi) &\text{ if } \zij = \shi.
\end{cases}
\end{equation}

Hence, if lane arrival probabilities are uniform, i.e. $p = p_j$, then
\begin{equation}
\Prob{\zi} = \prod_{j=1}^{Q-1} \Prob{\zij} = \Prob{\q} \text{ with } \q = h_i(\zi).
\end{equation}

Accordingly, if lane arrival probabilities are uniform, transition probabilities given by \eqref{eq:prob} and \eqref{eq:dprob} are equal for any states $\q,\q' \in \Siq$ and $\zi,\zip \in \Siz$ such that $\q = h_i(\zi)$ and $\q' = h_i(\zip)$, i.e. $\Prob{\zip|\zi} = \Prob{h_i(\zip) | h_i(\zi)}$. Hence, if the one-step-costs are such that $g_z(\zi) = g_q(h_i(\zi))$, the solutions of the systems of equations \eqref{eq:wq2} and \eqref{eq:WTZ} verify $\Wiz = W_q(\vi,\q)$ for any pair of states $\zi \in \Siz$ and $\q \in \Siq$ such that $\q = h_i(\zi)$.
\end{proof}
\label{propmap}
\end{prop}
	
Proposition \ref{propmap} establishes a correspondence between the proposed queue- and lane-based models for determining the expected waiting times of users and shows that under uniform lane arrival probabilities and one-step-costs, both models are equivalent. We next illustrate the lane-based model and the mapping $\q = h_i(\zi)$.

\begin{example}
\label{ex:lane}
We continue Example \ref{ex:queue} and focus on an intersection with $Q=3$ lanes. To determine the expected waiting of a state $\zi \in \Siz$ requires solving the systems of equations \eqref{eq:WTZ} with $\left\vert\Tiz\right\vert=2^{Q-1} = 4$ terminal states and $\left\vert\Siz\right\vert - \left\vert\Tiz\right\vert = 3^{Q-1} - 4 = 5$ non-terminal states. The state space with all 9 states corresponding to MC of the lane-based model is depicted in Figure \ref{fig:mc_lane}.

As in Example \ref{ex:queue}, we examine the queue-based state $\q = (1,0)$. The set of lane-based states $\zi \in \Siz$ which are antecedents of $\q$ via the mapping $\q = h_i(\zi)$ is, denoted $\Siz(h_i^{-1}((1,0)))$ is:
\[\Siz\left(h_i^{-1}((1,0))\right) = \left\{\left(\shi,\slo\right), \left(\slo,\shi\right)\right\}.\]

Further, we assume a uniform distribution of users' delay costs in the range $\vlb = $\$5/hour and $\vub = $\$10/hour, i.e. $\mathcal{U}(5,10)$, and a unit one-step-cost of $g_z(\zi) = 1$ s. According to Proposition \ref{propmap}, if $p_j = p$ for all lanes $j$ of the intersection and if $g_z(\zi) = g_q(h_i(\zi))$ for all states $\zi$, then $W_q(\vi,(1,0)) = W_z(\vi,\zi)$ for all $\zi \in \Siz(h_i^{-1}((1,0)))$.

Denote $\zi=(j_1,j_2)$, with $j_1$ and $j_2$ the lanes other than that of user $i$. We consider non-uniform lane arrival probabilities $p_i = 1/3, p_{j_1} = 1/2$ and $p_{j_2} = 1/6$, and as in Example \ref{ex:queue} we examine the case of user $i$ with a value of time of \$7/hour bidding $\vi=0.19$~\textcent.

\begin{itemize}
\item For state $\zi = \left(\shi,\slo\right)$, the expected waiting time is $W_z\left(\vi,\left(\shi,\slo\right)\right) = 1.43$ s.
\item For state $\zi = \left(\slo,\shi\right)$, the expected waiting time is $W_z\left(\vi,\left(\slo,\shi\right)\right) = 1.11$ s.
\end{itemize}
	
Observe that taking the average lane arrival probability $p = (1/3 + 1/2 + 1/6)/3 = 1/3$ and using the queue-based model, we obtain an expected waiting time of $1.25$ s (see Example \ref{ex:queue}), which differs from the expected arrival times obtained using either states $\zi \in \Siz\left(h_i^{-1}((1,0))\right)$ of the lane-based model corresponding to the queue-based state $\q=(1,0)$. This highlights the impact of lane-specific arrival probabilities onto users' expected waiting time.
	
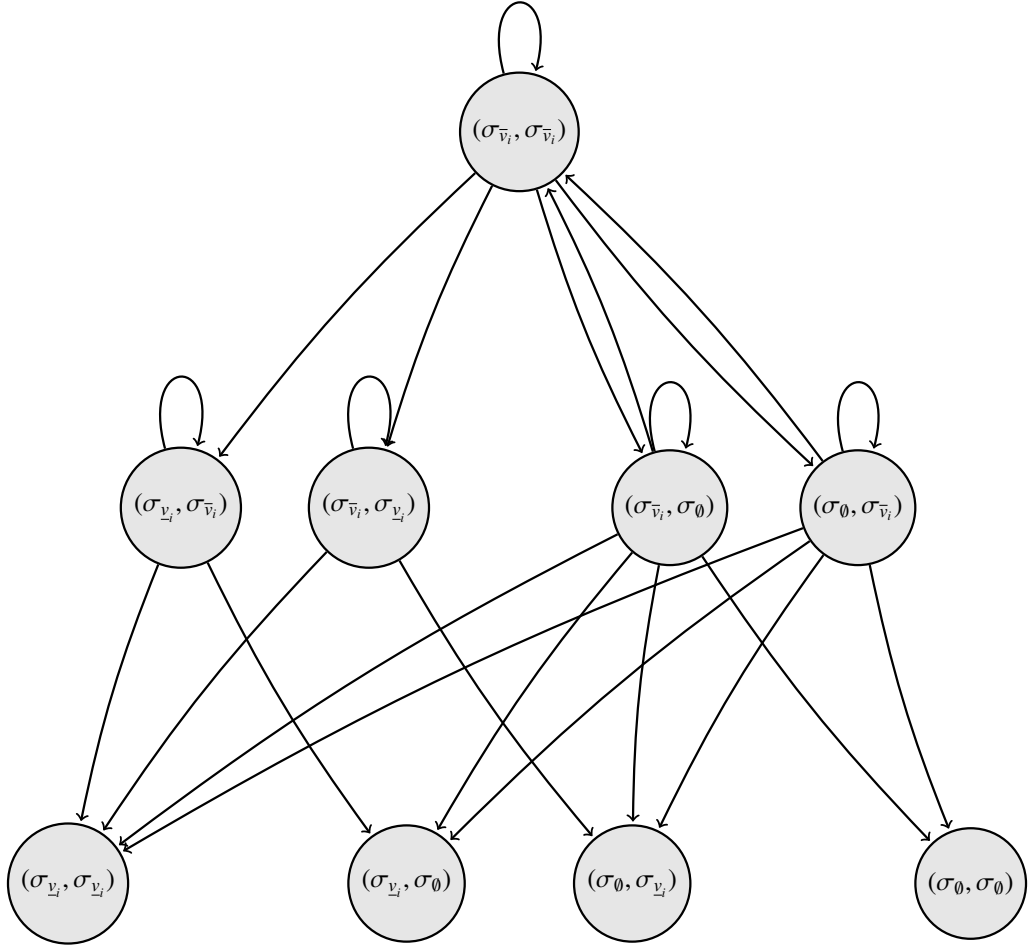
\begin{figure}
\centering
\begin{tikzpicture}
\tikzset{node style/.style={state, 
		minimum width=1cm,
		line width=0.3mm,
		fill=gray!20!white}}
\node[node style] at (7, 0) (hh) {$(\shi,\shi)$};
\node[node style] at (2.5, -5) (lh) {$(\slo,\shi)$};
\node[node style] at (5, -5) (hl) {$(\shi,\slo)$};	
\node[node style] at (9, -5) (he) {$(\shi,\se)$};	
\node[node style] at (11.5, -5) (eh) {$(\se,\shi)$};
\node[node style] at (1, -10) (ll) {$(\slo,\slo)$};
\node[node style] at (5.5, -10) (le) {$(\slo,\se)$};
\node[node style] at (8.5, -10) (el) {$(\se,\slo)$};	
\node[node style] at (13, -10) (ee) {$(\se,\se)$};
\draw[every loop,auto=right,line width=0.3mm]
(hh)     edge[loop above]            node {} (hh)
(lh)     edge[loop above]            node {} (lh)
(hl)     edge[loop above]            node {} (hl)
(he)     edge[loop above]            node {} (he)
(eh)     edge[loop above]            node {} (eh)			
(hh)     edge[bend right=5]         node {} (lh)
(hh)     edge[bend right=5]         node {} (hl)
(hh)     edge[bend right=5]         node {} (he)		
(hh)     edge[bend right=5]         node {} (eh)
(he)     edge[bend right=5]         node {} (hh)		
(eh)     edge[bend right=5]         node {} (hh)
(lh)     edge[bend right=5]         node {} (ll)
(hl)     edge[bend right=5]         node {} (ll)	
(he)     edge[bend right=5]         node {} (ll)	
(eh)     edge[bend right=5]         node {} (ll)		
(lh)     edge[bend right=5]         node {} (le)	
(he)     edge[bend right=5]         node {} (le)	
(eh)     edge[bend right=5]         node {} (le)	
(hl)     edge[bend right=5]         node {} (el)	
(he)     edge[bend right=5]         node {} (el)	
(eh)     edge[bend right=5]         node {} (el)
(he)     edge[bend right=5]         node {} (ee)	
(eh)     edge[bend right=5]         node {} (ee);	
\end{tikzpicture}
\caption{Illustration of the MC diagram for the lane-based model corresponding to 3-lane intersection ($Q=3$). The number of states is $|\Siz|=3^{Q-1}=9$. The states are written in the form $\left(z_{ij_1},z_{ij_2}\right)$ where $j_1$ and $j_2$ are the indices of lanes other than that of user $i$.}
\label{fig:mc_lane}
\end{figure}
\end{example}

We next present a payment mechanism that can be used with both of these models to determine incentive-compatible payments.
	
\subsection{Payment mechanism}
\label{payment}

The proposed payment mechanism can be used with both queue- and lane-based models for determining the expected waiting time of users. Hence, in this section, we simply denote $\Wvi$ the expected waiting time of user $i$ bidding $\vi$. This expected waiting time can be replaced by $\Wiq$ for the queue-based model, or by $\Wiz$ for the lane-based model. To determine incentive-compatible user payments, we calculate the expected marginal cost user $i$ imposes on other users. For this, we first examine the period of time over which an extra user in the lane of user $i$ is expected to have an impact on other users. 

The busy period w.r.t. user $i$ is the expected time required to clear the queues if $i$ declares a minimal delay cost, i.e. $\vi = \vlb$. Observe that if $\vi=\vlb$, then $\qli=0$, since no user can bid lower than $\vlb$ i.e. $F(\vlb)=0$. We measure the impact of user $i$ onto other users throughout the remaining busy period, i.e. the duration between the start of service of user $i$ if $i$ declares $\vi$, and the start of service of $i$ if $i$ had declared a minimal delay cost. Hence, the remaining busy period w.r.t. user $i$ bidding $\vi$ denoted $\DWvi$ can be defined as:
\begin{equation}
\DWvi \equiv \Wvlb - \Wvi.
\label{eq:rmb1}
\end{equation}

To determine the marginal impact of user $i$ on other users, we can split the remaining busy period w.r.t. user $i$ into two components: the expected portion of $\DWvi$ which affects users who arrived \textit{before} user $i$; and the expected portion of $\DWvi$ which affects users who are expected to arrive \textit{after} user $i$. We denote $\Bvi$ and $\Avi$ these quantities, respectively, and intuitively we define
\begin{equation}
\DWvi \equiv \Bvi + \Avi.
\label{eq:rmb2}
\end{equation}

Calculating $\Bvi$ and $\Avi$ correctly is not trivial, especially the latter since the exact number of users arriving after $i$ that will be delayed by user $i$ is unknown. We next propose a method to determine $\Bvi$, then use Eq. \eqref{eq:rmb2} together with Eq. \eqref{eq:rmb1} to obtain $\Avi$.

To determine the marginal delay caused by user $i$ onto users already in the pricing queue, we need only to consider those $\qli$ users which bid lower than $i$ since others users are not delayed by user $i$. Let $\Vli$ be the set of all users bidding lower than $i$ ordered by increasing declared delay costs, i.e. $\Vli = \{j_1, j_2, \ldots, j_{\qli}\}$ with $v_{j_1} < v_{j_2} < \ldots < v_{j_{\qli}} < \vi$. To determine the expected portion of the remaining busy period w.r.t. user $i$ which affects user $j \in \Vli$, we take the perspective of an extra user bidding $\vj$ in the lane of user $i$. The expected period of time during which this extra user affects user $j$ is the difference between the waiting time of the extra user if the extra user is bypassed by user $j$ and that if the extra user bypasses user $j$ in the pricing queue. We abuse notation and denote $W_i(\vj : \zij = \shi)$ and $W_i(\vj : \zij = \slo)$ these expected waiting times, respectively. 

We propose the following formula for $\Bvi$:
\begin{equation}
\Bvi = \sum_{k = 1}^{\qli} W_i\left(v_{j_k} : z_{ij_k} = \shi\right) - W_i\left(v_{j_k} : z_{ij_k}  = \slo\right).
\label{eq:b}
\end{equation}

The definitional relationship linking the time periods $\Bvi$ and $\DWvi$, i.e. Eq. \eqref{eq:rmb2}, implicitly assumes that $\Bvi \leq \DWvi$. However, it is not obvious that the right-hand side of Eq. \eqref{eq:b} is never greater than $\DWvi$. Since this is necessary for the correctness of the proposed formula for the time period $\Bvi$, i.e. Eq. \eqref{eq:b}, we prove this relationship in Proposition \ref{rmbb}.

\begin{prop}
For any user $i$ bidding $\vi$,
\begin{equation}
\sum_{k = 1}^{\qli} W_i(v_{j_k} : z_{ij_k} = \shi) - W_i\left(v_{j_k} : z_{ij_k}  = \slo\right) \leq \DWvi.
\label{eq:rmb3}
\end{equation}

\begin{proof}
\begin{align}
\sum_{k = 1}^{\qli}  W_i\left(v_{j_k} : z_{ij_k} = \shi\right) - W_i\left(v_{j_k} : z_{ij_k} = \slo\right) =& W_i\left(v_{j_1} : z_{ij_1} = \shi\right) - W_i\left(v_{j_1} : z_{ij_1} = \slo\right) + W_i(v_{j_2} : z_{ij_2} = \shi) \nonumber \\
&- W_i\left(v_{j_2} : z_{ij_2}  = \slo\right) + \ldots +  W_i\left(v_{j_{\qli}} : z_{ij_{\qli}} = \shi\right) - W_i\left(v_{j_{\qli}} : z_{ij_{\qli}} = \slo\right) \label{eq:bexpand}
\end{align}

Recall that $v_{j_k} < v_{j_{k+1}} < \vi$ for any $k = 1 \ldots \qli-1$. Observe that the state corresponding to the extra user bidding $v_{j_k}$ and bypassing user $j_k$ is equivalent to the state of the extra user bidding $v_{j_{k+1}}$ and being bypassed by user $j_{k+1}$, i.e. these states have identical number and location of lower/higher-bidders and empty lanes w.r.t. the perspective of the extra user. Since $v_{j_k} < v_{j_{k+1}}$, the expected waiting time of the extra user bidding $v_{j_{k+1}}$ and being bypassed by user $j_{k+1}$ is greater than the expected waiting time of the extra user bidding $v_{j_k}$ and bypassing user $j_k$, i.e.
\[
W_i\left(v_{j_{k+1}} : z_{ij_{k+1}} = \shi\right) - W_i\left(v_{j_k} : z_{ij_k}  = \slo\right) \leq 0 \qquad \forall k = 1 \ldots \qli-1.
\]
 
Omitting these negative terms from Eq. \eqref{eq:bexpand} yields:
\begin{equation}
\sum_{k = 1}^{\qli} W_i\left(v_{j_k} : z_{ij_k} = \shi\right) - W_i\left(v_{j_k} : z_{ij_k}  = \slo\right) \leq W_i\left(v_{j_{1}} : z_{ij_{1}}  = \shi\right) - W_i\left(v_{j_{\qli}} : z_{ij_{\qli}}  = \slo\right)
\label{eq:bvileq}
\end{equation}

The state corresponding to the extra user bidding $v_{j_{\qli}}$ and bypassing user $j_{\qli}$ is equivalent to the state from the perspective of user $i$. Thus, since $v_{j_{\qli}} < \vi$, $W_i(v_{j_{\qli}} : z_{ij_{\qli}}  = \slo) > \Wvi$. Finally, $\Wvlb$ is an upper bound on the waiting time for any user on the lane of user $i$, thus $W_i(v_{j_{1}} : z_{ij_{1}}  = \shi) < \Wvlb$. Substituting in Eq. \eqref{eq:bvileq} gives
\[
\sum_{k = 1}^{\qli} W_i(v_{j_k} : z_{ij_k} = \shi) - W_i(v_{j_k} : z_{ij_k}  = \slo)  \leq \Wvlb - \Wvi  = \DWvi.
\]
\end{proof}
\label{rmbb}
\end{prop}

Proposition \ref{rmbb} together with Eq. \eqref{eq:rmb2} ensure that the expected portion of the remaining busy period corresponding to future arrivals is nonnegative, i.e. $\DWvi - \Bvi = \Avi \geq 0$. Using Eq. \eqref{eq:b}, the expected marginal delay cost that user $i$ imposes on users who arrived before her can be determined as:
\begin{equation}
\MBvi = \sum_{k = 1}^{\qli} \big(W_i(v_{j_k} : z_{ij_k} = \shi) - W_i(v_{j_k} : z_{ij_k}  = \slo)\big) v_{j_k}.
\label{eq:mb}
\end{equation}

To determine the expected marginal delay cost that user $i$ imposes on future arrivals, denoted $\MAvi$, we use the result of \cite{dolan1978incentive} who observed that this marginal delay cost can be calculated by integrating $\frac{d\Av}{dv} v$ over the bid range $[\vlb,\vi)$. Further, the author observed that the state is constant for any bid comprised between two consecutive bids of users in the pricing queue. Thus, for any segment corresponding to a pair of consecutive bids in the sequence $\{\vlb,v_{j_1}, \ldots, v_{j_{\qli-1}},v_{j_{\qli}},\vi\}$, the expected marginal delay cost imposed by user $i$ onto users bidding in this segment can be calculated by integrating $\frac{d\Av}{dv} v$ over the corresponding domain. Specifically, let $v_{j_0} = \vlb$ and let $v_{j_{\qli+1}} = \vi$. Let $D(k) = (v_{j_k},v_{j_{k+1}})$ be the domain corresponding to the bid segment $(v_{j_m},v_{j_n})$ for $m < n$ such that $j_m, j_n \in \Vli$. Accordingly, the expected marginal delay cost imposed by user $i$ on future arrivals can be determined as:
\begin{equation}
\MAvi = \sum_{k=0}^{\qli}\int_{D(k)} \frac{d \Av}{dv} v dv = \sum_{k=0}^{\qli}\int_{D(k)} \frac{-dW_i(v)}{dv} v dv.
\label{eq:ma}
\end{equation} 

The expected marginal delay cost imposed by user $i$ declaring $\vi$ on other users is thus:
\begin{equation}
\MCvi = \MBvi + \MAvi.
\label{eq:mc}
\end{equation}

We next give the main result. 

\begin{theorem}
If the user objective function is $\min_{\vi} \vit \Wvi + \Pvi$ and users are serviced in order of decreasing declared delay costs, the payment $\Pvi = \MCvi$ is incentive-compatible.

\begin{proof}
The proof follows the same logic as that of \cite{dolan1978incentive}. We next recall its main steps and detail elements which are specific to our mechanism. Consider a pricing queue of size $Q$ and a user with true delay cost $\vit$ declaring $\vit + \delta$ with $\delta > 0$. Increasing the declared delay cost may reduce waiting time in two ways: i) bypassing a user in the queue, and, ii) having fewer future arrivals bypass the bidder. 

For i) the reduction in expected waiting time is $B_i(\vit + \delta) - B_i(\vit) = W_i(v_{j_k} : z_{ij_k} = \shi) - W_i(v_{j_k} : z_{ij_k} = \slo) \geq 0$ for some user $k$ such that $v_{j_k} \in [\vit,\vit+\delta)$. This reduction is valued $\vit(B(\vit + \delta) - B(\vit))$ but the added cost is $v_{j_k}(B(\vit + \delta) - B(\vit))$ with $v_{j_k} \geq \vit$. 

For ii) the reduction in expected waiting time is $A_i(\vit + \delta) - A_i(\vit)$ which is valued $\vit(A_i(\vit + \delta) - A_i(\vit))$. However, the induced expected marginal cost is $MA_i(\vit + \delta) - MA_i(\vit) = \int_{\vit}^{\vit + \delta} \frac{d\Av}{dv} v dv \geq \int_{\vit}^{\vit + \delta} \frac{d\Av}{dv} \vit dv = \vit(A_i(\vit + \delta) - A_i(\vit))$.
A similar reasoning can be applied for users declaring a delay cost lower than their true delay cost. 
\end{proof}
\label{theo}
\end{theorem}

Theorem \ref{theo} proves that the expected marginal delay cost as defined by Eq. \eqref{eq:mc} represents an incentive-compatible user payment in the dynamic sense, i.e. by taking into account the expected impact of users on future arrivals. This provides a basis to implement the proposed queue- and lane-based mechanisms in an online fashion. Upon reaching the front of their lane, users are asked to declare their delay cost $\vi$ and receive a corresponding payment $\Pvi$ which ensures strategyproof user behavior in the long run. 

\begin{example}
\label{ex:payment}

We illustrate the payment mechanism by continuing Examples \ref{ex:queue} and \ref{ex:lane}. We consider a 3-lane intersection with a unit one-step cost of 1 s, i.e. $\gq = \gz = 1$, which represents the time it takes for a vehicle to traverse the intersection. We assume that users' delay cost follow a uniform distribution $\mathcal{U}(5,10)$ and examine the case of a reference user $i$ with a value of time of \$7/hour which corresponds to a bid of $\vi=0.19$~\textcent. We consider that the intersection is in the queue-based state $\q = (0,1)$ corresponding to one lane occupied by a higher-bidder and one lane occupied by a lower-bidder.

\begin{itemize}
\item If lane arrival probabilities are uniform, i.e. $p_j = p = 1/3$ for all lane $j$ of the intersection, using the queue-based model we obtain an expected waiting time of $1.25$ s (see Example \ref{ex:queue}). Using the payment mechanism, we determine a payment $MC_i(\vi) = 0.45$~\textcent\ and which corresponds to a generalized user cost of $C_i(\vi) = 0.69$ \textcent.

\item If lane arrival probabilities are non-uniform with $p_i = 1/3$, $p_{j_1} = 1/2$ and $p_{j_2} = 1/6$ there are two lane-based states which correspond to the queue-based state $(1,0)$: $\zi = (\shi,\slo)$ and $\zi = (\slo,\shi)$ (see Example \ref{ex:lane}). For state $(\shi,\slo)$, the expected waiting time is $1.43$ s and the payment is $MC_i(\vi) = 0.43$~\textcent\ which corresponds to a generalized user cost of $C_i(\vi) = 0.71$~\textcent. For state $(\slo,\shi)$, the expected waiting time is $1.11$ s and the payment is $MC_i(\vi) = 0.48$~\textcent\ which corresponds to a generalized user cost of $C_i(\vi) = 0.70$~\textcent.
\end{itemize}

Further details on the calculations are summarized in Table \ref{tab:pay}.

\begin{table}[!b]
\begin{tabular}{llllllllll}
\toprule
\textbf{Model} & \textbf{State} & $\Wvi$ & $\Wvlb$ & $\Bvi$ & $\Avi$ & $MB_i(\vi)$ & $MA_i(\vi)$ & $MC_i(\vi)$ & $C_i(\vi)$\\
\midrule
Queue & $(1,0)$ & 1.25 & 4.12 & 1.93 & 0.94 & 0.32 & 0.13 & 0.45 & 0.69 \\
Lane & $(\shi,\slo)$ & 1.43 & 4.19 & 1.65 & 1.11 & 0.27	 & 0.15 & 0.43 & 0.71 \\
Lane & $(\slo,\shi)$ & 1.11 & 4.19 & 2.16 & 0.92 & 0.36 & 0.12 & 0.48 & 0.70 \\
\bottomrule
\end{tabular}

\caption{Summary of calculations for an intersection with $Q=3$ lanes and using the queue- and lane-based mechanisms with a unit one-step cost of $\gq = \gz = 1$ s. The queue-based state $\q=(1,0)$ and its corresponding lane-based states $\zi = (\shi,\slo)$ and $\zi = (\slo,\shi)$ are illustrated. Users' delay cost are assumed to follow a uniform distribution $\mathcal{U}(5,10)$ and user $i$ is assumed to have a value of time of \$7/hour. Expected waiting times ($\Wvi$ and $\Wvlb$) and remaining busy period components ($\Bvi$ and $\Avi$) are reported in s. Expected marginal cost components ($MB_i(\vi)$, $MA_i(\vi)$ and $MC_i(\vi)$) and generalized costs ($C_i(\vi)$) are reported in \textcent.}
\label{tab:pay}
\end{table}
\end{example}

The proposed queue- and lane-based traffic intersection mechanisms are computationally efficient. In terms of computational resources, both mechanisms require the solution of Eqs. \eqref{eq:mb} and \eqref{eq:ma} to determine user payments. Eq. \eqref{eq:mb} requires the solution of multiple systems of linear equations (\eqref{eq:wq2} or \eqref{eq:WTZ} depending on the mechanism selected) which can be accomplished using standard linear algebra codes. To compute Eq. \eqref{eq:ma}, standard numerical integration and differentiation techniques can be used in combination with codes for systems of linear equations. As will be shown in our numerical experiments, both mechanisms have low average computation time per user and can thus be expected to be executed in real-time to determine incentive-compatible payments.

\section{Numerical experiments}
\label{num}

\begin{figure}[t]
	\centering
	\begin{tabular}{lr}
		\begin{subfigure}[b]{0.5\linewidth}
			\centering\includegraphics[width=0.9\linewidth]{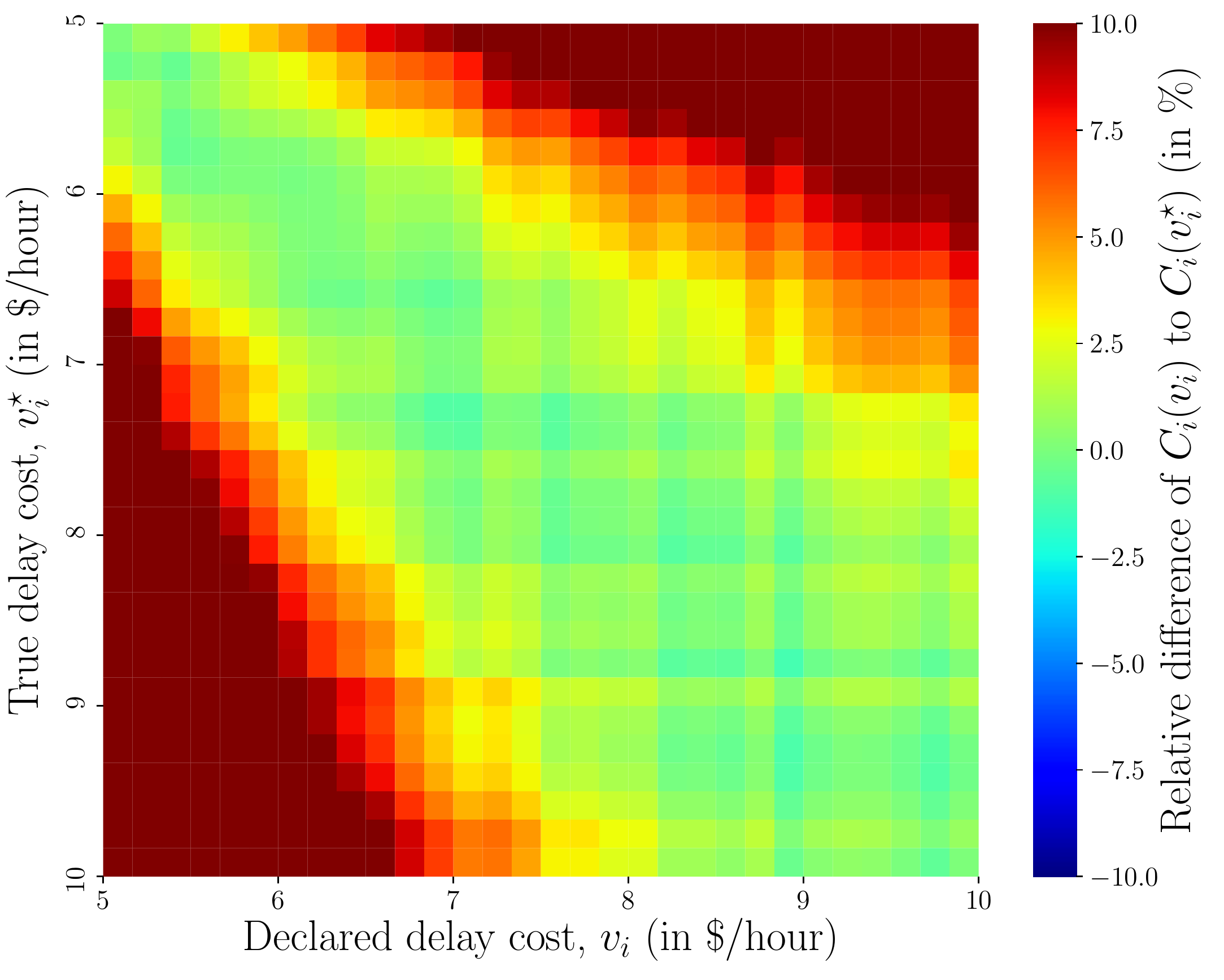}
			\caption{Queue-based mechanism.\label{fig:queue}}
		\end{subfigure}%
		\begin{subfigure}[b]{0.5\linewidth}
			\centering\includegraphics[width=0.9\linewidth]{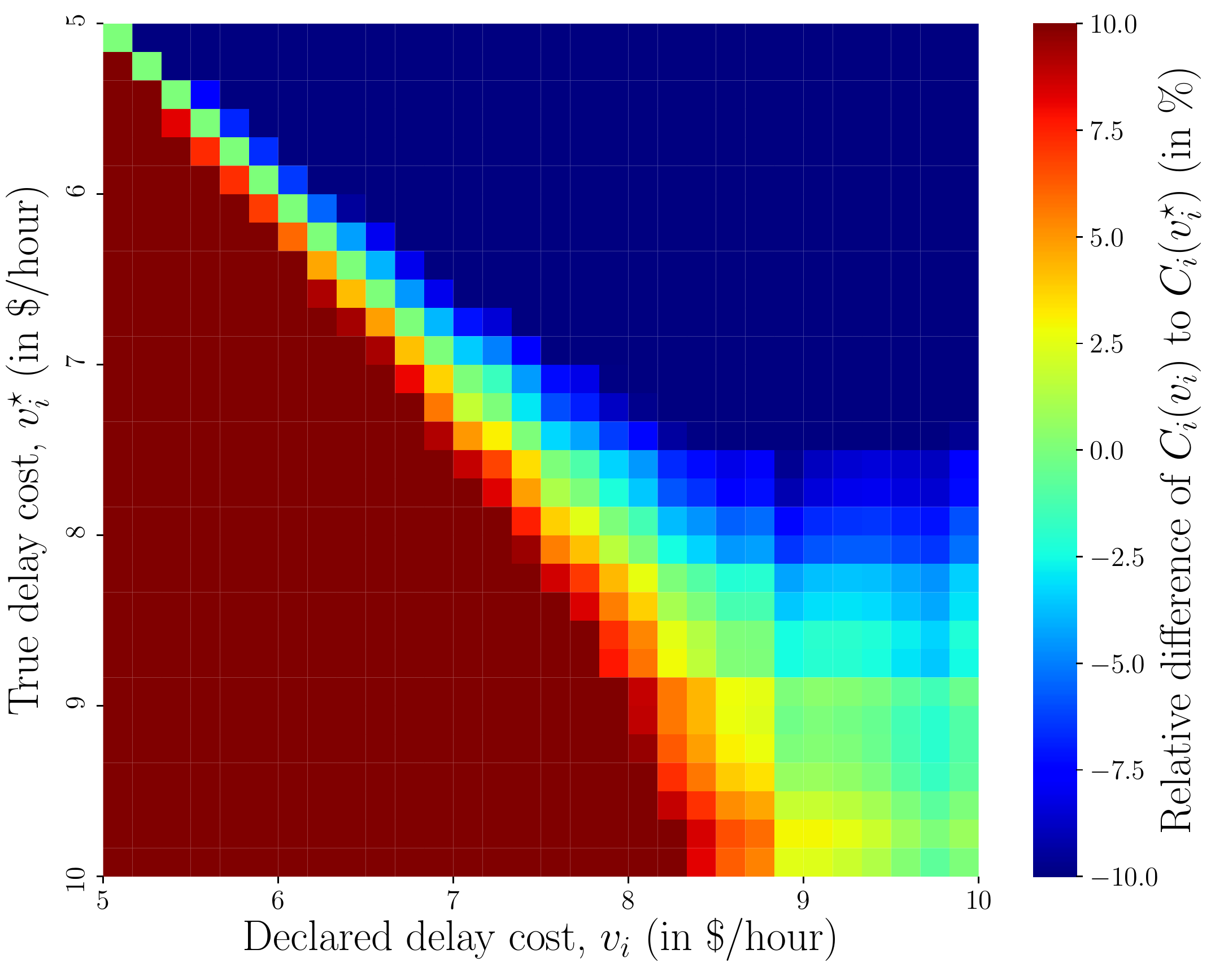}
			\caption{Static mechanism.\label{fig:static}}
		\end{subfigure}
	\end{tabular}
	\caption{Relative user generalized cost $\left(\Cvi - C_i\left(v_i^\star\right)\right)/C_i\left(v_i^\star\right)$. Results of a simulation with 1 million users with an arrival probability of $p=0.25$. Figure~\ref{fig:queue} shows relative user generalized costs using the queue-based online mechanism and Figure~\ref{fig:static} shows relative user generalized costs obtained using the static mechanism.\label{fig:qs}}
\end{figure}

Numerical simulations of the proposed online traffic intersection mechanisms are conducted to explore their behavior. 

We implement the proposed online mechanisms by simulating the arrival and departure of users at an intersection using a discrete time process. At every time period, random trials are conducted for each empty lane to simulate the stochastic arrival process of users. For all our experiments, we use a unit one-step cost, i.e. $\gq = \gz = 1$ for all states $\q \in \Siq$ and $\zi \in \Siz$. We set the time period duration to 1 s, i.e. representing a uniform user service time of 1 s. This value is chosed based on the geometric configuration of typical 4-lane intersection and average vehicle speeds \citep{levin2017conflict}. Users' delay costs are randomly drawn from a uniform distribution $\mathcal{U}(5,10)$, i.e. $\vlb=\$5$ /hour and $\vub=\$10$ /hour. These values are chosen based on a study on value of time for automated vehicles \citep{neeraj2018}. This delay cost range corresponds to user bids between 0.14~\textcent\ and 0.28~\textcent\ for a time period duration of 1 s which aims to represent users' willingness to pay for service priority at traffic intersections.

To quantify the value of using a dynamic auction compared to a static auction, we implement the static mechanism outlined by \citet{dolan1978incentive} for priority queueing systems. This static mechanism is a Vickrey-Clarke-Groves auction which is known to be incentive-compatible in the static sense, i.e. all participating users are assumed to be known when determining payments \citep{vickrey1961counterspeculation,clarke1971multipart,groves1975incentives}. The static mechanism utilizes the same priority queueing model than that of the proposed online mechanism, i.e. users in the pricing queue are sorted by decreasing declared delay costs, but differ in the determination of waiting time and user payments. In the static mechanism, the waiting time of user $i$ is only function of the position of user $i$ in the pricing queue, e.g. if user $i$ is third in the pricing queue, her waiting time is the sum of the two one-step costs for transitioning from the current state to the state where user $i$ served. The payment of user $i$ is sum of user valuations over lower-bidders in the pricing queue. This payment is incentive-compatible in the static sense \citep{dolan1978incentive}.

In each experiment, we simulate the service of one million users and report the average user behavior by segmenting users' delay cost. Specifically, we segment the delay cost range $[5,10]$ into 30 uniform bins and group all serviced users throughout the simulation into these bins. We then report average waiting times, payments and generalized user costs based on the average quantities obtained for each of these 30 bins. The simulations are implemented in Python and the linear systems of equations \eqref{eq:wq2} and \eqref{eq:WTZ} are solved using Numpy's linear algebra solver on a Windows 10 machine with 8 Gb of RAM and a CPU of 2.7 Ghz.\\ 

We first compare the behavior of the proposed traffic intersection auctions mechanisms on a four-lane intersection in Section \ref{ulanes} under uniform lane arrival probabilities. We then consider non-uniform lane arrival probabilities on a four-lane intersection in Section \ref{nulanes}. The computational performance of the proposed mechanisms is examined in Section \ref{runtime} for traffic intersections with a varying number of lanes.

\subsection{Uniform lane arrival probabilities}
\label{ulanes}

\begin{figure}
	\centering\includegraphics[width=0.45\linewidth]{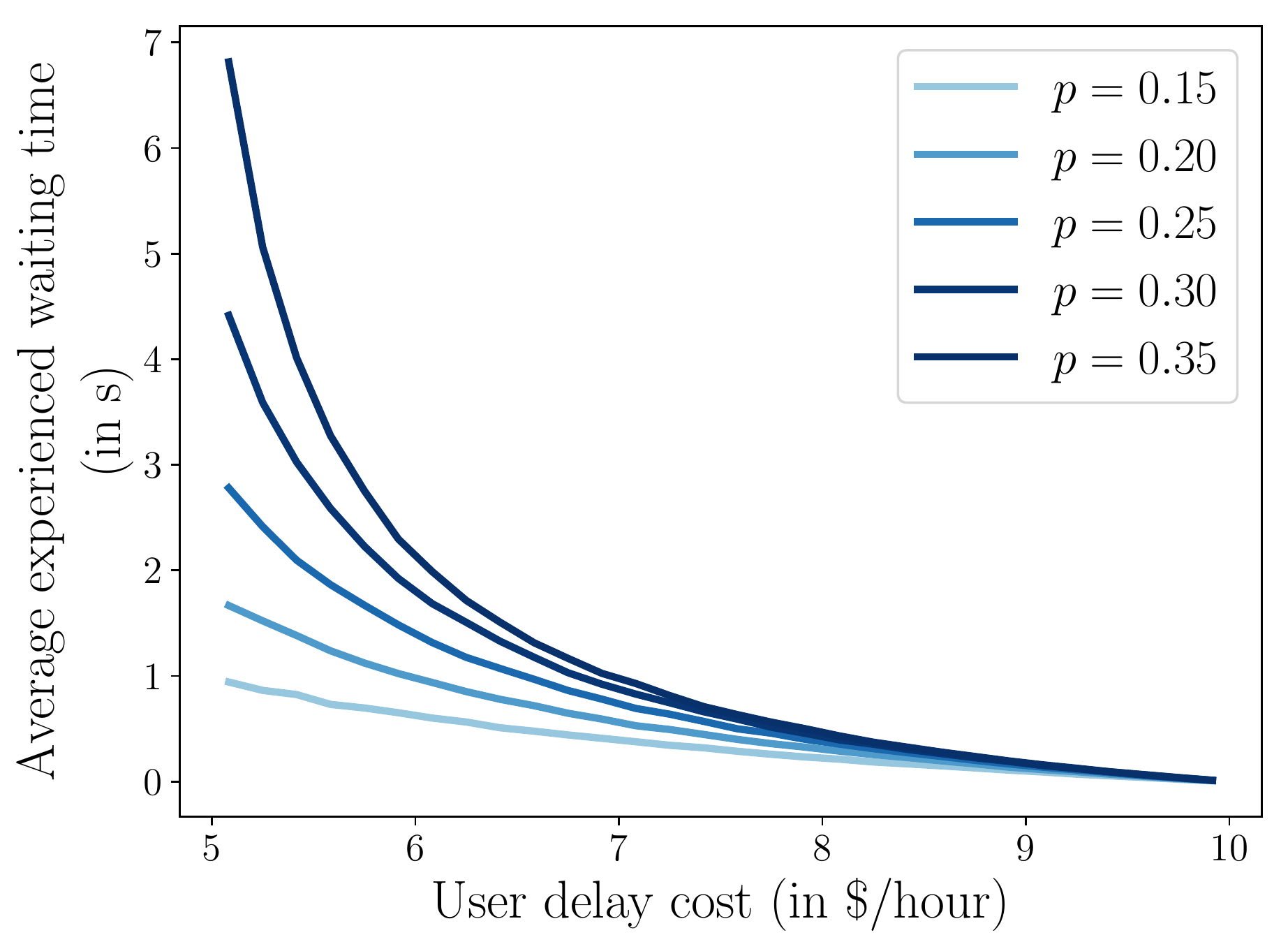}
	\caption{Average experienced waiting time of users based on their delay cost. Results of five simulations with 1 million users each with a varying arrival probability $p$.\label{fig:awtqs}}
\end{figure}

In this section, we consider uniform lane arrival probabilities, i.e. $p = p_j$ for all lanes $j$ of the intersection. In this configuration, as shown by Proposition \ref{propmap}, the queue- and lane-based mechanisms yield identical outcomes. Hence, in this section, we solely compare the proposed queue-based mechanism with the static mechanism.\\

We first set the expected arrival rate of users at the intersection to 1. Since the arrival rate is uniform across all $Q=4$ lanes, this corresponds to an arrival probability of $p=0.25$. The results are reported in Figure \ref{fig:qs} which depicts the relative difference of $\Cvi$ to $C_i\left(v_i^\star\right)$ in percentage, which represents the difference between the generalized cost of user $i$ declaring a delay cost $\vi$ and the generalized cost of user declaring its true delay cost $v_i^\star$. Strictly positive values indicate that declaring $\vi$ results in a higher cost than declaring the user true delay cost $\vit$, whereas negative values indicate that misreporting delay cost $\vi$ achieves a lower cost than truthful reporting. For clarity, relative user generalized cost values are bounded within $\pm 10\%$, i.e. if $\left(\Cvi - C_i\left(v_i^\star\right)\right)/C_i\left(v_i^\star\right)$ is greater (resp. lower) than $10\%$ (resp. $-10\%$), this value is shown as $10\%$ (resp. $-10\%$). 

\begin{figure}
	\centering
	\begin{subfigure}[b]{0.49\linewidth}
		\centering\includegraphics[width=0.95\linewidth]{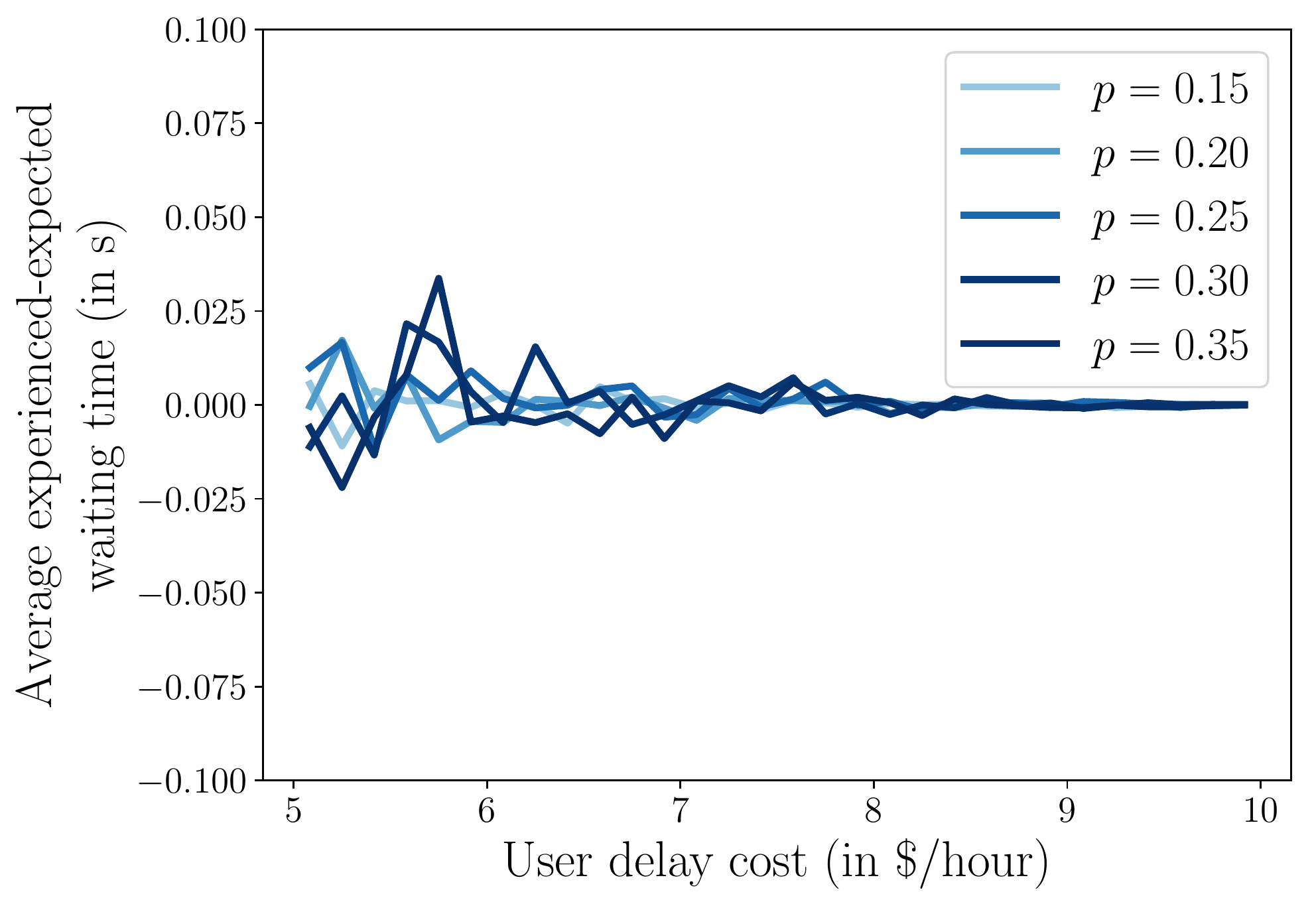}
		\caption{Queue-based mechanism.\label{fig:diffwtqueue}}
	\end{subfigure}
	\begin{subfigure}[b]{0.45\linewidth}
		\centering\includegraphics[width=0.95\linewidth]{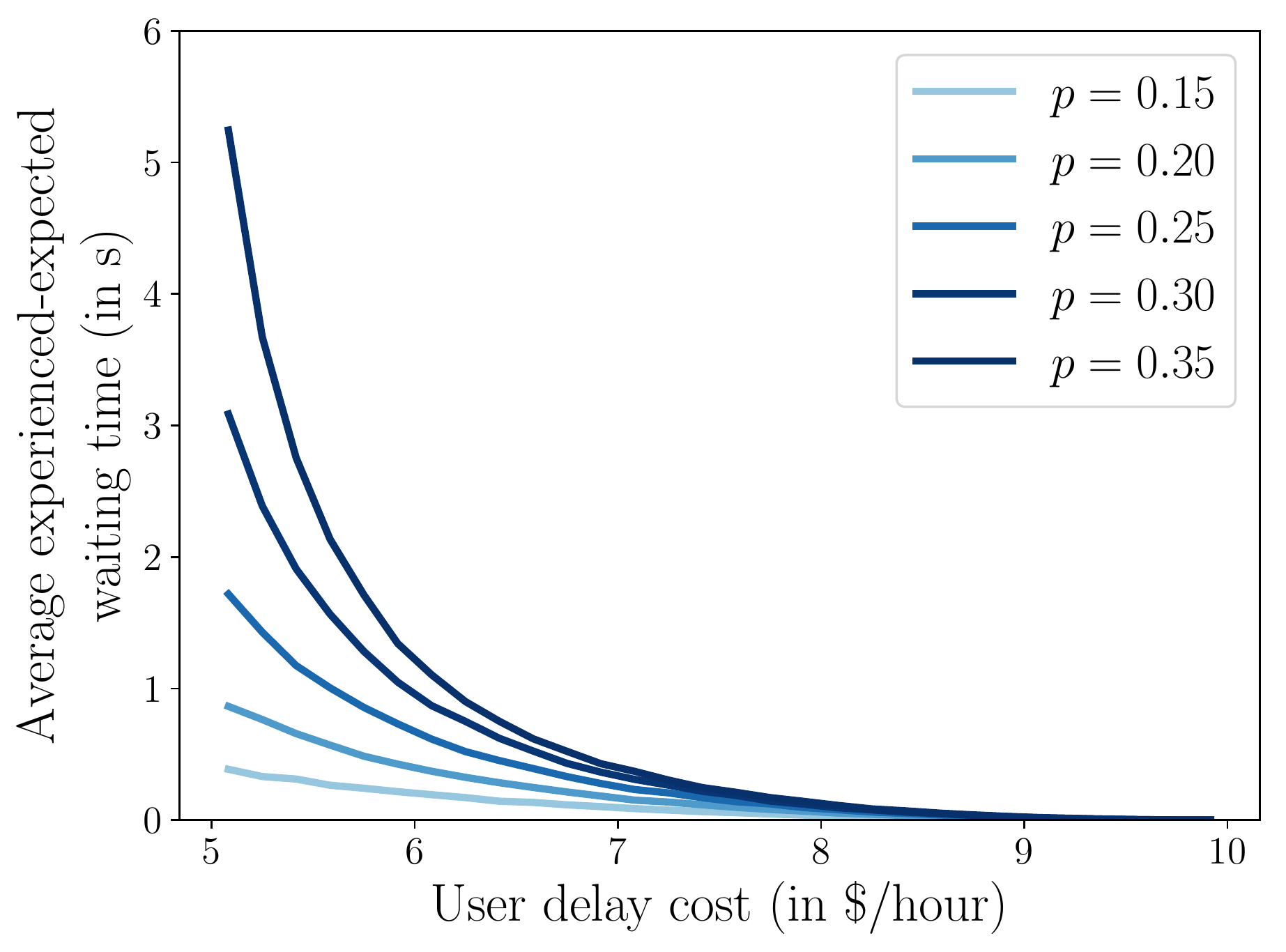}
		\caption{Static mechanism.\label{fig:diffwtstatic}}
	\end{subfigure}
	\caption{Results of simulations with 1 million users with a varying arrival probability of $p$. Figure~\ref{fig:diffwtqueue} shows the average experienced-expected waiting times obtained using queue-based online mechanism and Figure~\ref{fig:diffwtstatic} shows average experienced-expected waiting times obtained using the static mechanism.\label{fig:diffwt}}
\end{figure}

\begin{figure}
	\centering
	\begin{subfigure}[b]{0.45\linewidth}
		\centering\includegraphics[width=0.95\linewidth]{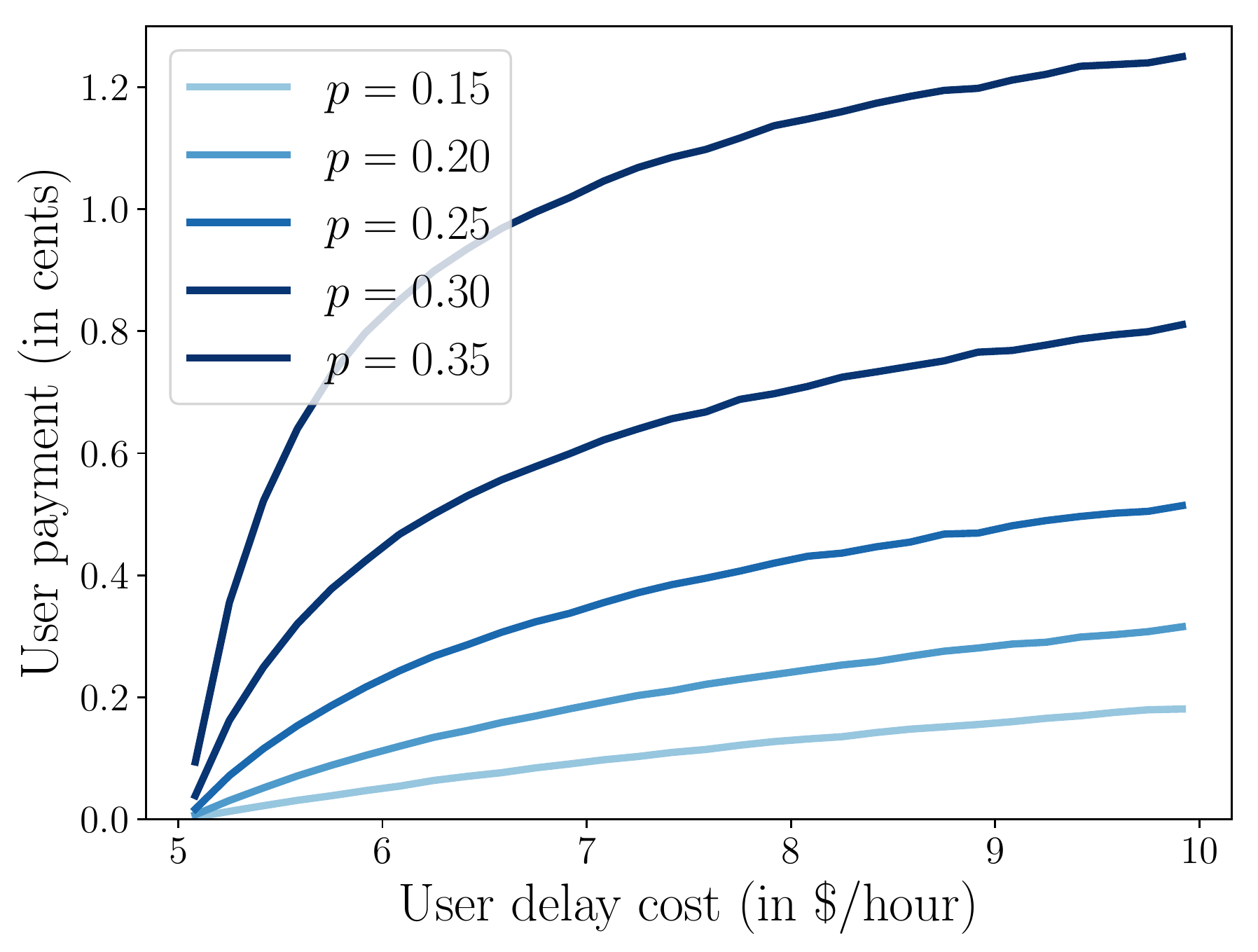}
		\caption{Queue-based mechanism.\label{fig:payqueue}}
	\end{subfigure}
	\begin{subfigure}[b]{0.45\linewidth}
		\centering\includegraphics[width=0.95\linewidth]{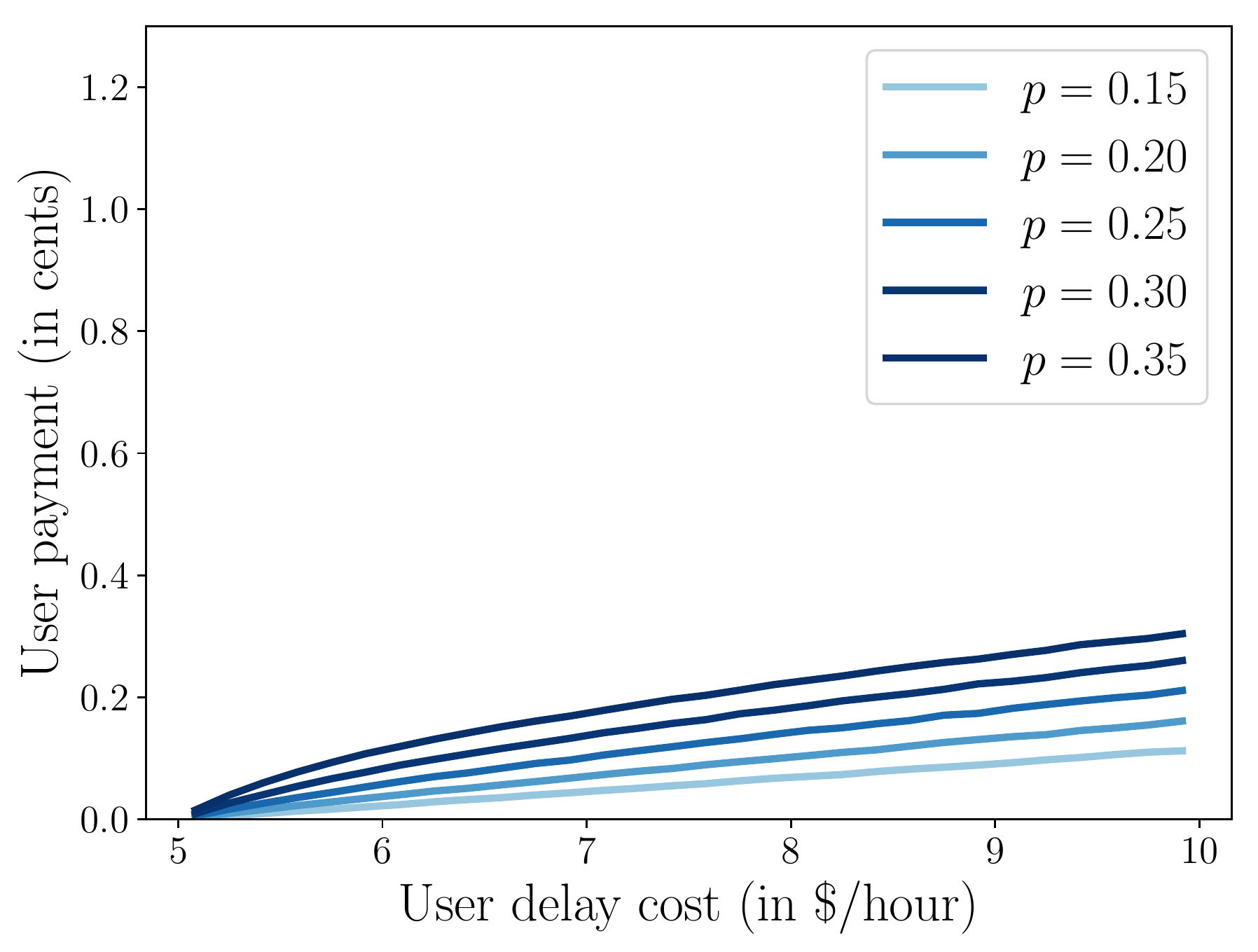}
		\caption{Static mechanism.\label{fig:paystatic}}
	\end{subfigure}
	\caption{Results of simulations with 1 million users with a varying arrival probability of $p$. Figure~\ref{fig:payqueue} shows user payments obtained using the queue-based online mechanism and Figure~\ref{fig:paystatic} shows user payments obtained using the static mechanism.\label{fig:payqs}}
\end{figure}

The relative user generalized cost obtained via the online queue-based mechanism is shown to be, on average, incentive-compatible almost everywhere, as highlighted by positive relative generalized user costs. In turn, using the static mechanism, a user misreporting her delay cost by declaring a higher delay cost can achieve a lower generalized cost than truthfully reporting her delay cost, as highlighted by the top-right negative relative user generalized costs. This highlights the benefits of online mechanisms over static approaches for traffic intersection auctions. 

We next examine the impact of varying the arrival probability. We run simulations from $p = 0.15$, which corresponds to an expected arrival rate of 0.6, to $p = 0.35$, which corresponds to an expected arrival rate of 1.4, in steps of $0.05$. The results of these simulations are reported in Figures~\ref{fig:awtqs}--Figure~\ref{fig:payqs}. Figure~\ref{fig:awtqs} depicts the average experienced waiting time of users based on their delay cost for varying arrival probability $p$. Since in both dynamic and static mechanisms, users are prioritized based on their declared delay cost, the experienced, i.e. simulated, waiting time is identical in both mechanisms. As expected, experienced waiting times decrease with the delay cost. The decrease rate is near-linear for delay costs greater than \$7/hour. For lower delay costs and high arrival probabilities, the decrease rate is exponential, which highlights the potentially long waiting times incurred by low-bidding users in a congested regime, i.e. $p > 0.25$. The decrease rate of experienced waiting time is lower in stable regimes, i.e. $p < 0.25$ and exhibits a more linear decay rate. 

To quantify the accuracy of the proposed MC models for estimating expected waiting times, we report the difference of experienced minus expected waiting times for the queue-based and the static mechanisms in  Figure~\ref{fig:diffwt}. For clarity y-axis scales are different in Figure~\ref{fig:diffwtqueue} and Figure~\ref{fig:diffwtstatic} representing the queue-based and static mechanisms, respectively. Figure~\ref{fig:diffwtqueue} shows that the average expected waiting times obtained via the MC queue-based model are very close to the average experienced waiting times, with maximum deviations approximately 0.05 time units reported for low-bidding users. In contrast, Figure~\ref{fig:diffwtstatic} shows that the static mechanism almost systematically underestimates the average experienced waiting times.  Further, the loss of quality increases with the arrival probability at the intersection. These trends are consequences of the static mechanism not accounting for the probability of future arrivals. Figure~\ref{fig:payqs} shows the average payments of users under both queue-based and static mechanisms. User payments increase with delay cost and are comparatively larger in the queue-based mechanism (Figure~\ref{fig:payqueue}) than in the static mechanism (Figure~\ref{fig:paystatic}). This emphasizes the potential impact of future arrivals which are neglected in static mechanisms. The queue-based mechanism also exhibits more concave-shaped payments with regards to user delay cost compared to the static mechanism which is more linear. We also find that the curvature of the payment curves increase with the arrival probability, highlighting the impact of the demand regime onto the payment mechanisms.

\subsection{Non-uniform lane arrival probabilities}
\label{nulanes}

In this section, we consider non-uniform lane arrival probabilities $p_j$ all lanes $j$ of a four-lane intersection. We first examine the proposed mechanisms in the light of incentive-compatibility for a specific lane arrival probability configuration before comparing the impact of varying lane arrival probabilities on users waiting time and payments.

\begin{figure}
	\centering
	\begin{subfigure}[b]{0.7\linewidth}
	\centering\includegraphics[width=1.0\linewidth]{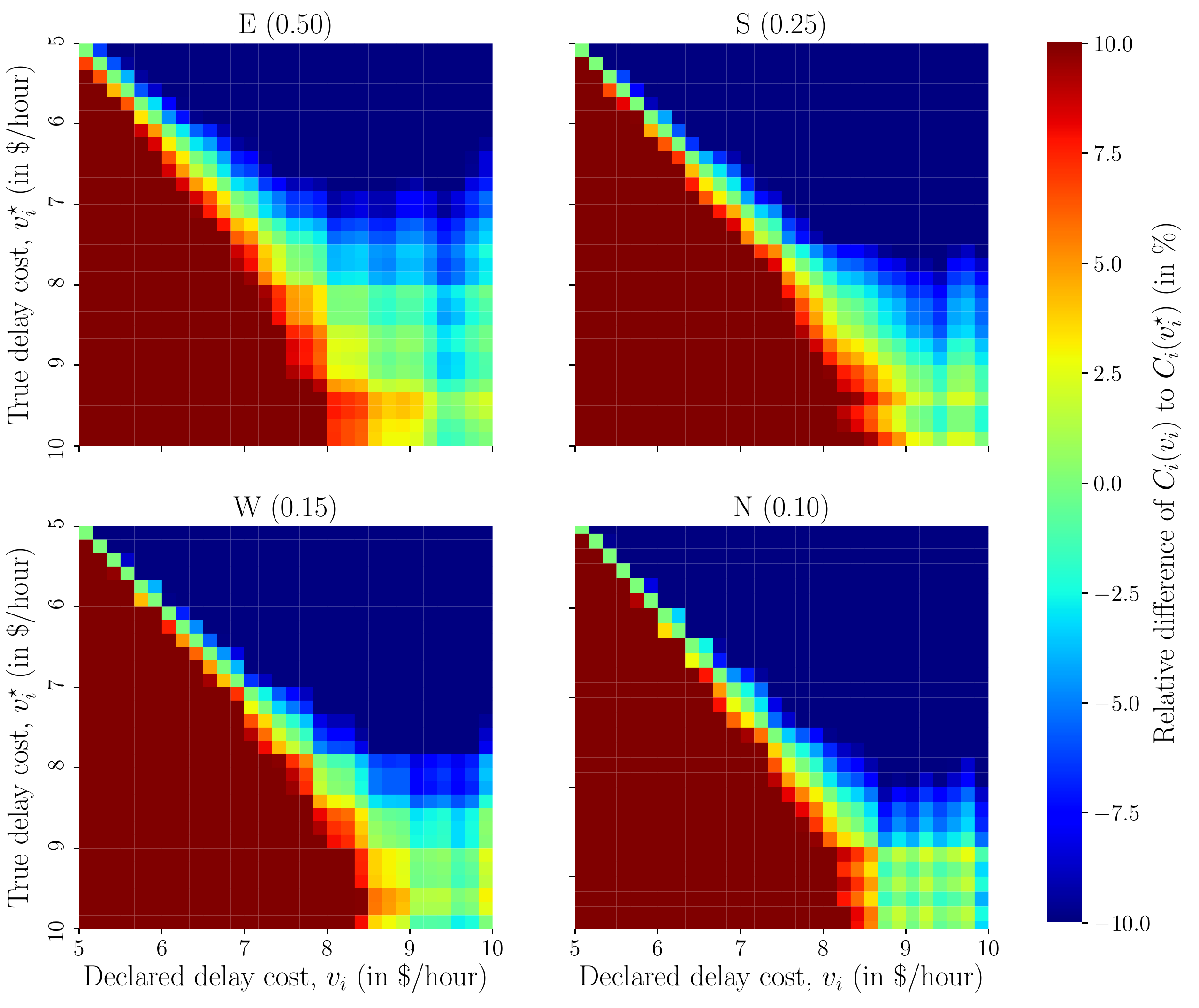}
	\caption{Static mechanism.\label{fig:heatmapls}}
\end{subfigure}\\	
	\begin{subfigure}[b]{0.7\linewidth}
		\centering\includegraphics[width=1.0\linewidth]{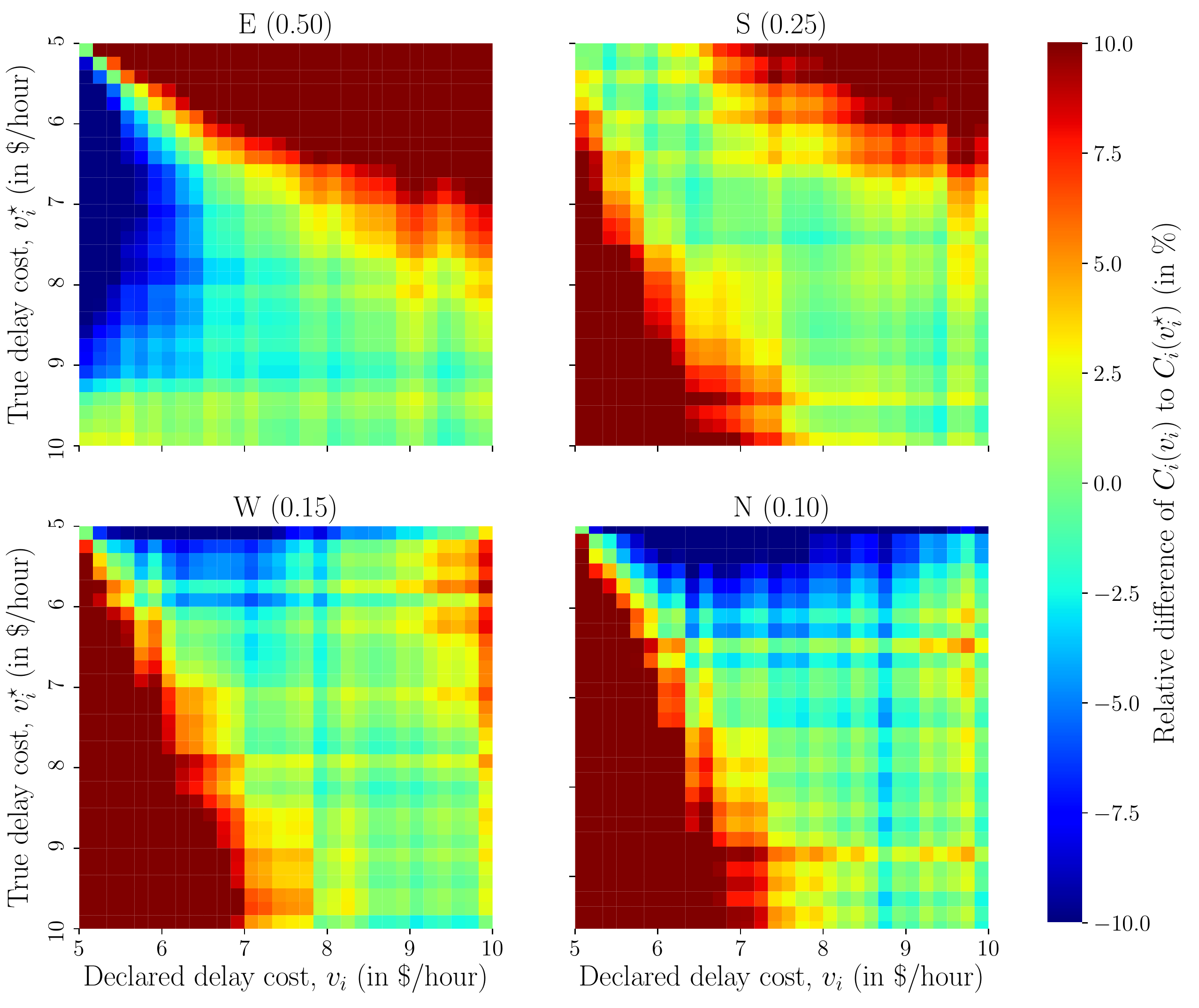}
		\caption{Queue-based mechanism.\label{fig:heatmaplq}}
	\end{subfigure}
\end{figure}
\begin{figure}\ContinuedFloat
	\begin{subfigure}[b]{0.7\linewidth}
		\centering\includegraphics[width=1.0\linewidth]{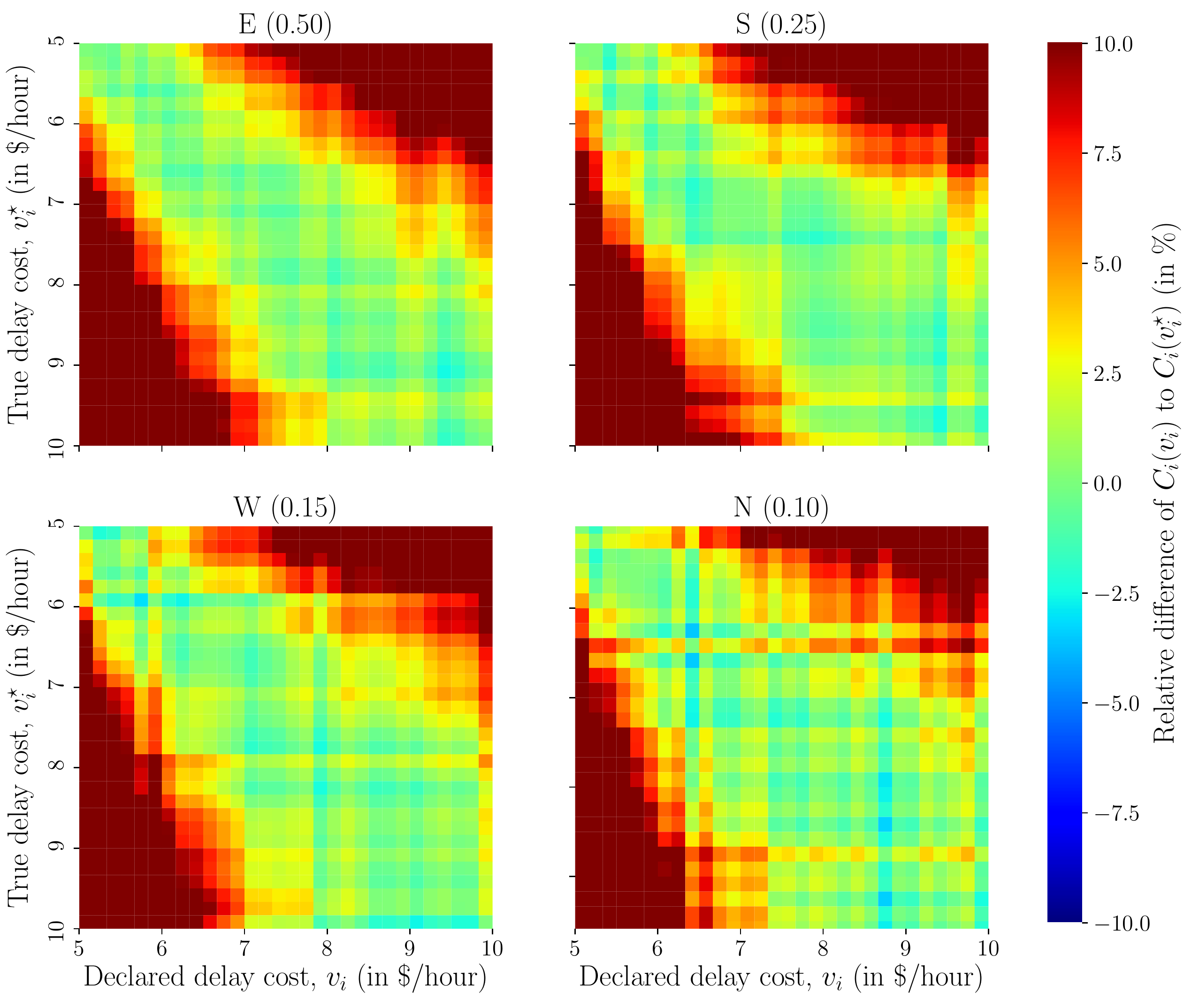}
		\caption{Lane-based mechanism.\label{fig:heatmapll}}
	\end{subfigure}
	\caption{Relative user generalized cost $\left(\Cvi - C_i\left(v_i^\star\right)\right)/C_i\left(v_i^\star\right)$. Results of a simulation with 1 million users with lane-based arrival probabilities of $0.50$, $0.25$, $0.15$ and $0.10$. Figure~\ref{fig:heatmapls} illustrates the static mechanism; Figure~\ref{fig:heatmaplq} illustrates the queue-based mechanism and Figure~\ref{fig:heatmapll} illustrates the lane-based mechanism.\label{fig:heatmaplane}}
\end{figure}

We use cardinal directions, i.e. East (E), South (S), West (W) and North (N) to refer to all $Q=4$ lanes of the intersection and assign lane arrival probabilities of $0.50$, $0.25$, $0.15$ and $0.10$, respectively. For the queue-based mechanism, we use the average lane arrival probability, i.e. $p = \sum_{j=1}^{Q} p_j = 0.25$. We report lane-based heatmaps of the relative difference in terms of relative user generalized cost $\left(\Cvi - C_i\left(v_i^\star\right)\right)/C_i\left(v_i^\star\right)$ for the static, queue- and lane-based mechanisms in Figure \ref{fig:heatmaplane}. Figure \ref{fig:heatmapls} shows the outcome of the simulation using the static mechanism. We find that the static mechanism does not promote truthful user behavior for all four lanes of the intersection, i.e. regardless of the lane arrival probability the mechanism is not incentive-compatible, and on all lanes users may reduce their generalized cost in the long run by over-reporting their delay cost. The outcome of the same simulation obtained using the queue-based mechanism is summarized in Figure \ref{fig:heatmaplq}. The outcome of these simulation highlight that using the average lane arrival probability in the queue-based mechanism fails to yields incentive-compatible outcome for the East, West and North lanes which have a lane arrival probability different that $p=0.25$. Only users on the South lane with $p_{S} = 0.25 = p$ receive incentive-compatible payments. In turn, Figure \ref{fig:heatmapll} illustrates that the lane-based mechanism is incentive-compatible for each lane of the intersection.

\begin{figure}
	\centering
	\begin{tabular}{cc}
		\begin{subfigure}{0.35\linewidth}
			\centering\includegraphics[width=1.0\linewidth]{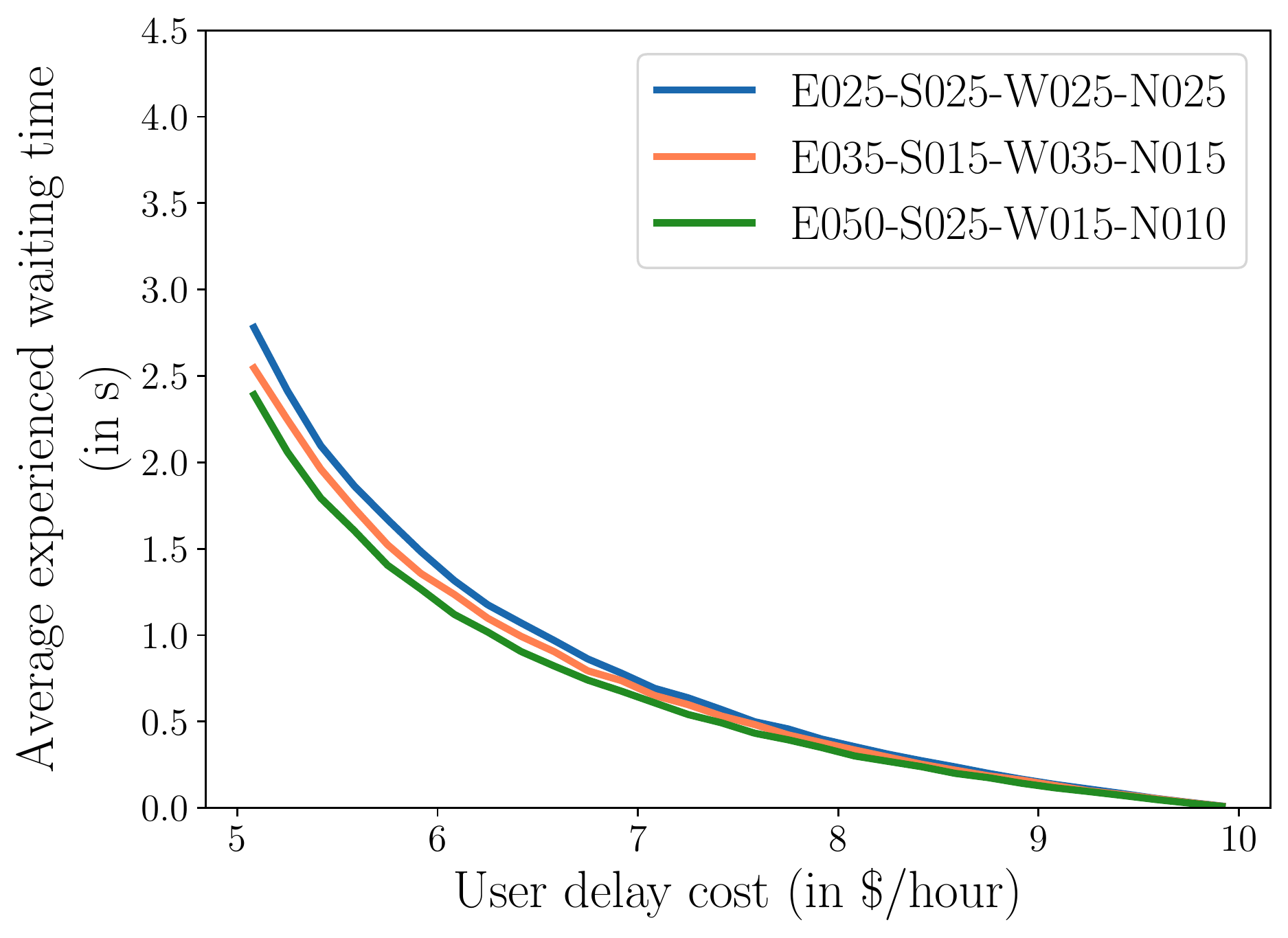}
			\caption{Average experienced waiting time over all four lanes. \label{fig:awtavglanes}}
		\end{subfigure} &
		\begin{subfigure}{0.35\linewidth}
			\centering\includegraphics[width=1.0\linewidth]{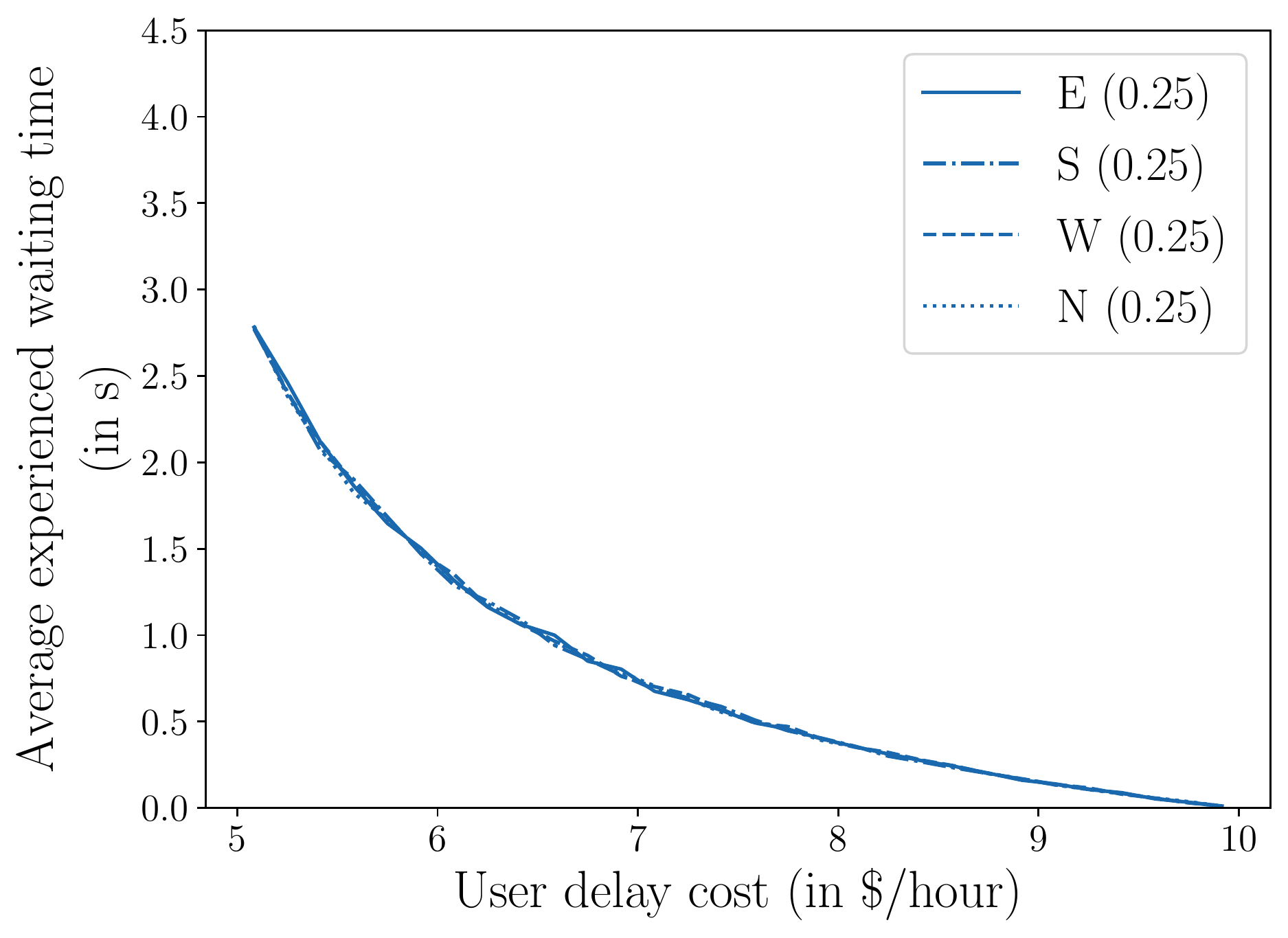}
			\caption{Lane-based average experienced waiting time for E025-S025-W025-N025.\label{fig:awt025}}
		\end{subfigure} \\
		\begin{subfigure}{0.35\linewidth}
			\centering\includegraphics[width=1.0\linewidth]{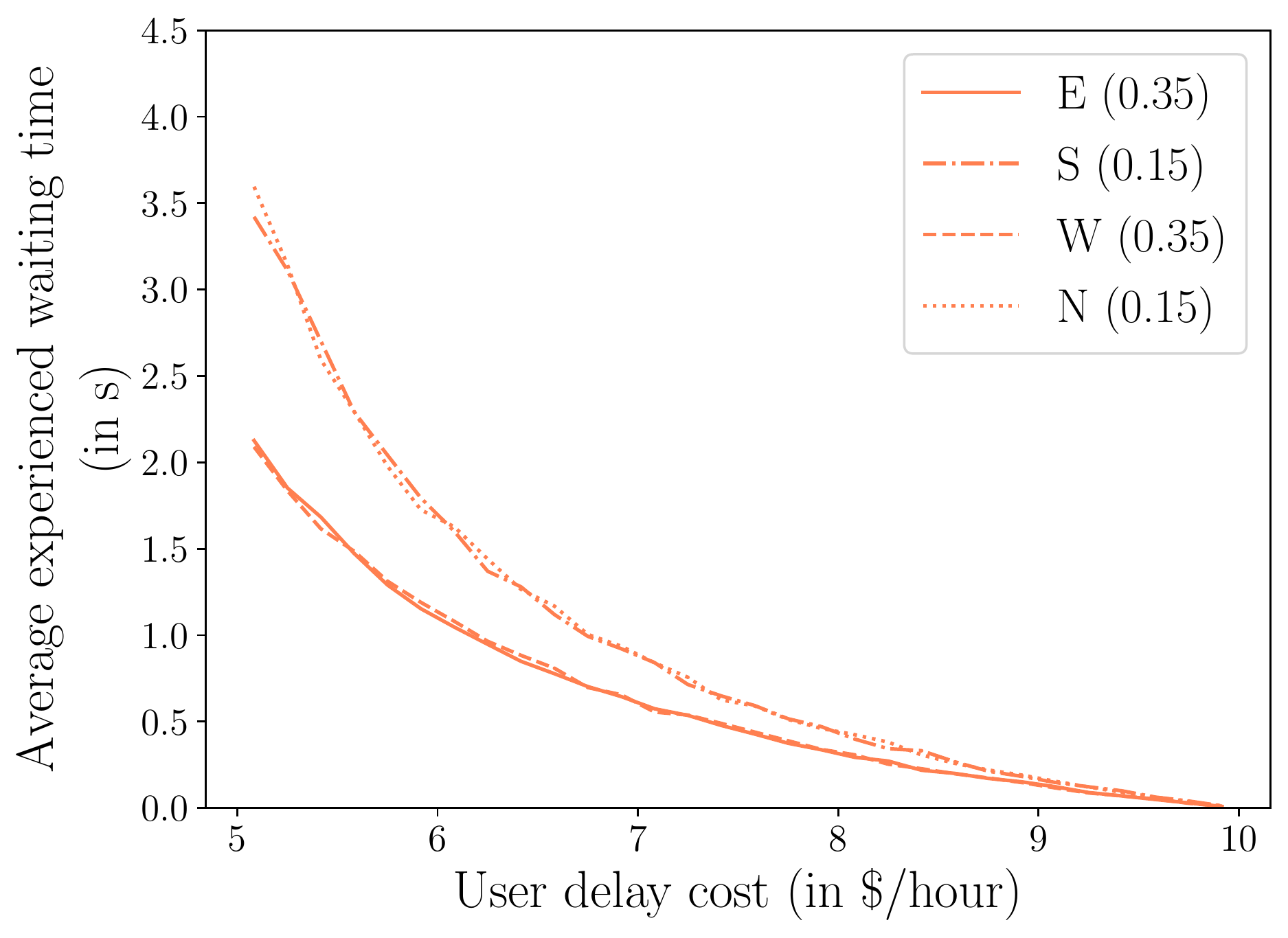}
			\caption{Lane-based average experienced waiting time for E035-S015-W035-N015.\label{fig:awt035}}
		\end{subfigure} &
		\begin{subfigure}{0.35\linewidth}
			\centering\includegraphics[width=1.0\linewidth]{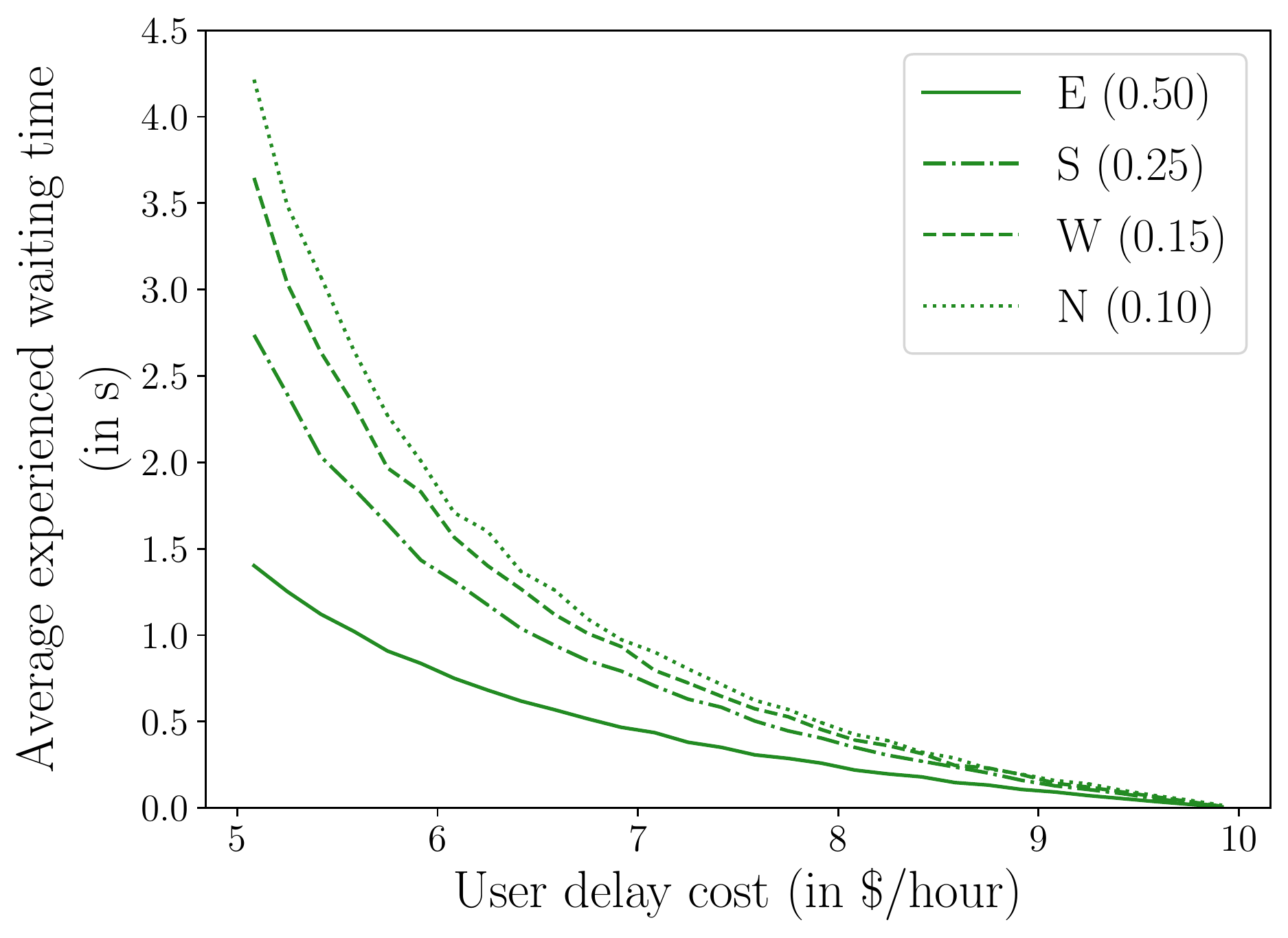}
			\caption{Lane-based average experienced waiting time for E050-S025-W015-N010.\label{fig:awt050}}
		\end{subfigure} \\
	\end{tabular}
	\caption{Average user experienced waiting times. Results of three simulations with varying lane-based arrival probabilities $p_j$, $j\in\{\text{E,S,W,N}\}$. Each simulation consists of 1 million users. \label{fig:awtlane}}
\end{figure}

\begin{figure}[!h]
	\centering
	\begin{tabular}{cc}
		\begin{subfigure}{0.35\linewidth}
			\centering\includegraphics[width=1.0\linewidth]{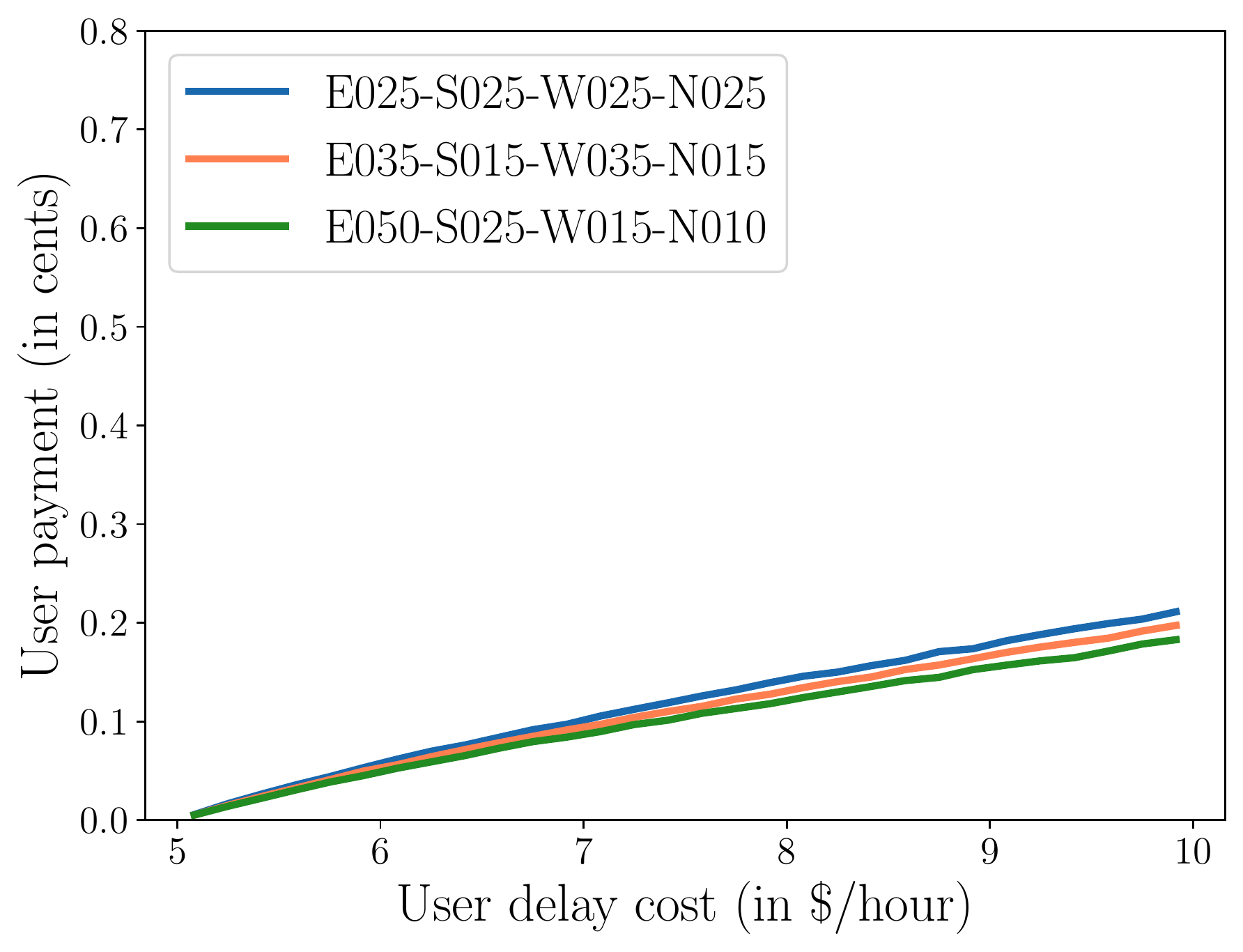}
			\caption{Average payment over all four lanes. \label{fig:payavglanes_static}}
		\end{subfigure} &
		\begin{subfigure}{0.35\linewidth}
			\centering\includegraphics[width=1.0\linewidth]{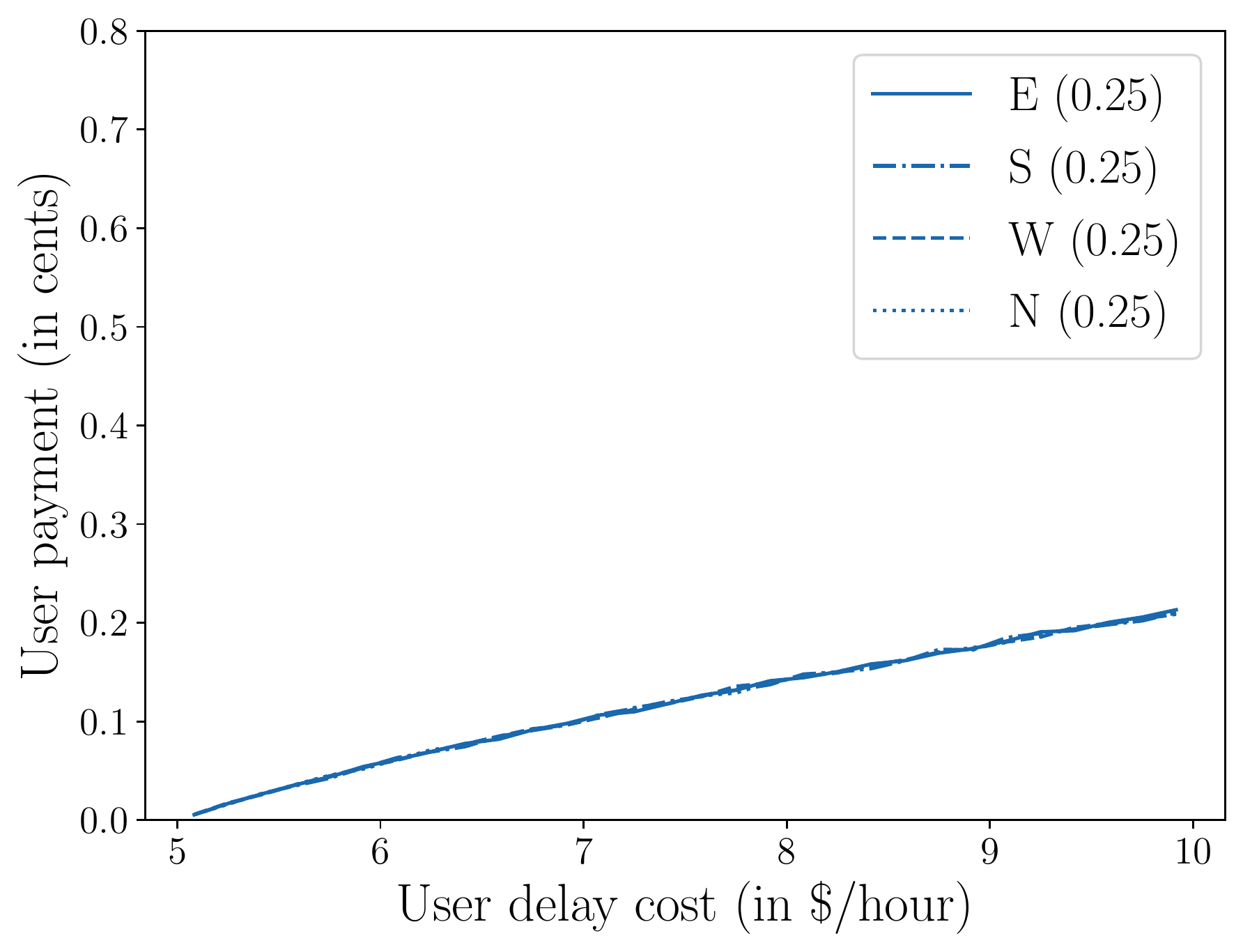}
			\caption{Lane-based payment for E025-S025-W025-N025.\label{fig:pay025_static}}
		\end{subfigure} \\
		\begin{subfigure}{0.35\linewidth}
			\centering\includegraphics[width=1.0\linewidth]{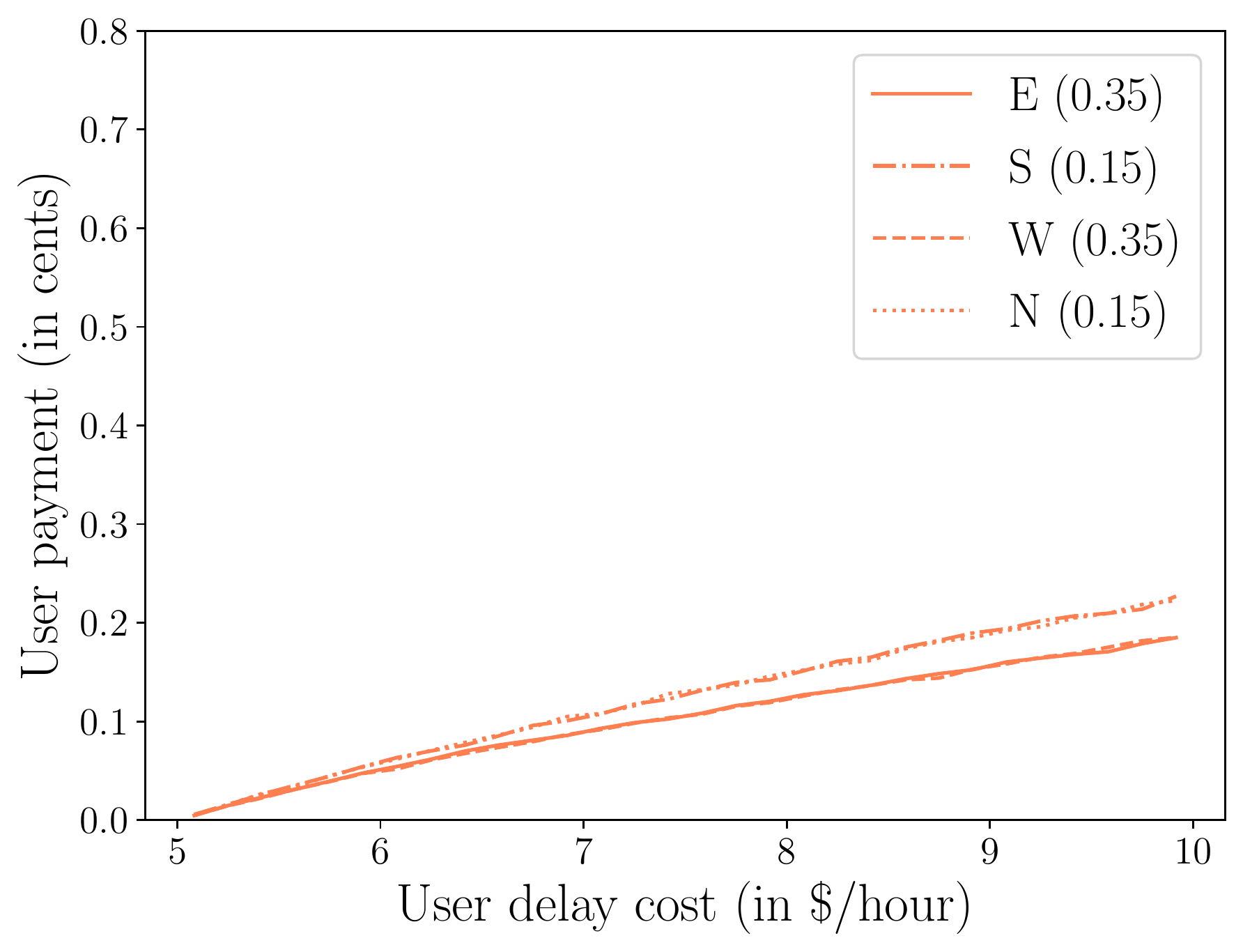}
			\caption{Lane-based payment for E035-S015-W035-N015.\label{fig:pay035_static}}
		\end{subfigure} &
		\begin{subfigure}{0.35\linewidth}
			\centering\includegraphics[width=1.0\linewidth]{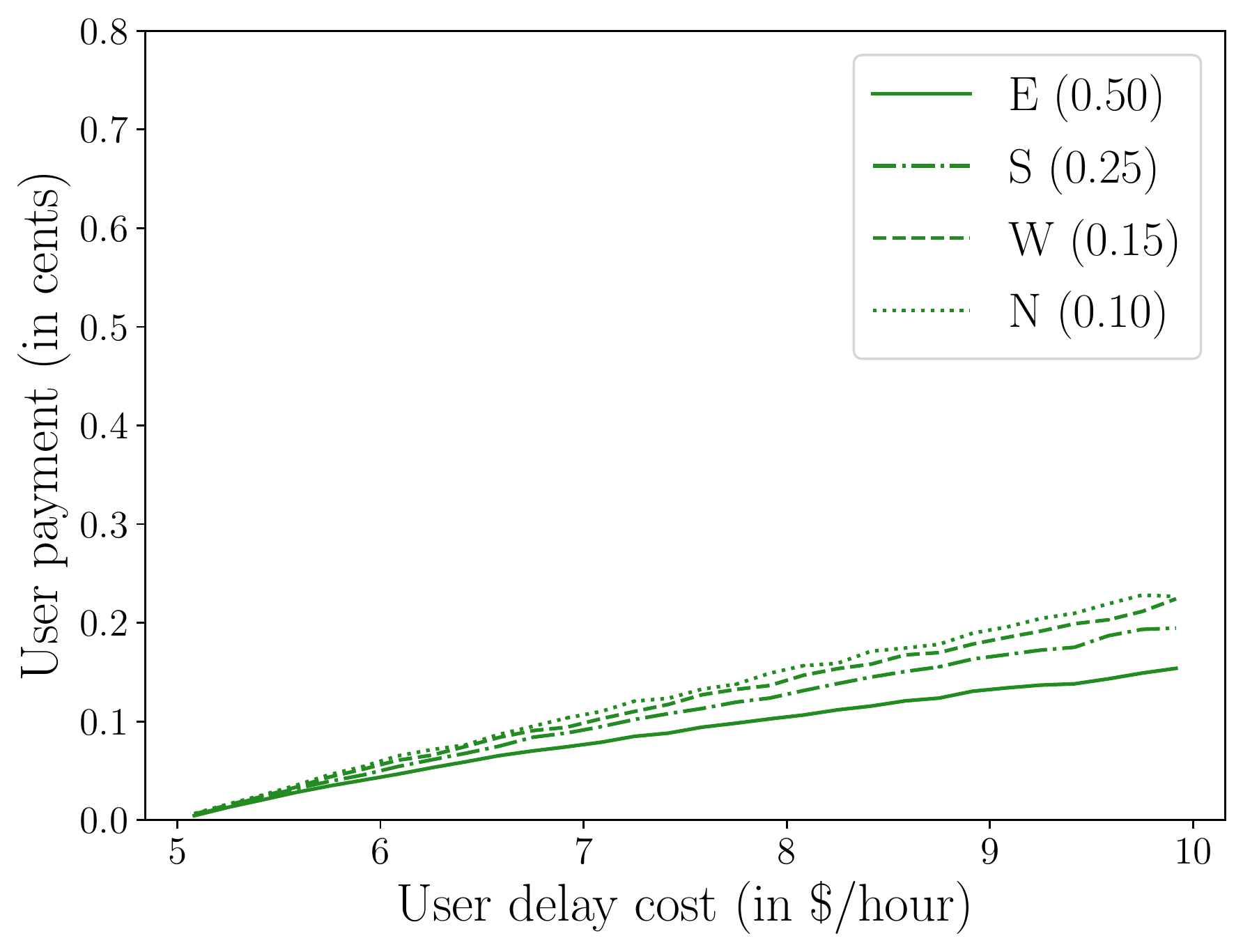}
			\caption{Lane-based payment for E050-S025-W015-N010.\label{fig:pay050_static}}
		\end{subfigure} \\
	\end{tabular}
	\caption{Average user payments using the static mechanism. Results of three simulations with varying lane-based arrival probabilities $p_j$, $j\in\{\text{E,S,W,N}\}$. Each simulation consists of 1 million users.\label{fig:paylane_static}}
\end{figure}

\begin{figure}[!h]
	\centering
	\begin{tabular}{cc}
		\begin{subfigure}{0.35\linewidth}
			\centering\includegraphics[width=1.0\linewidth]{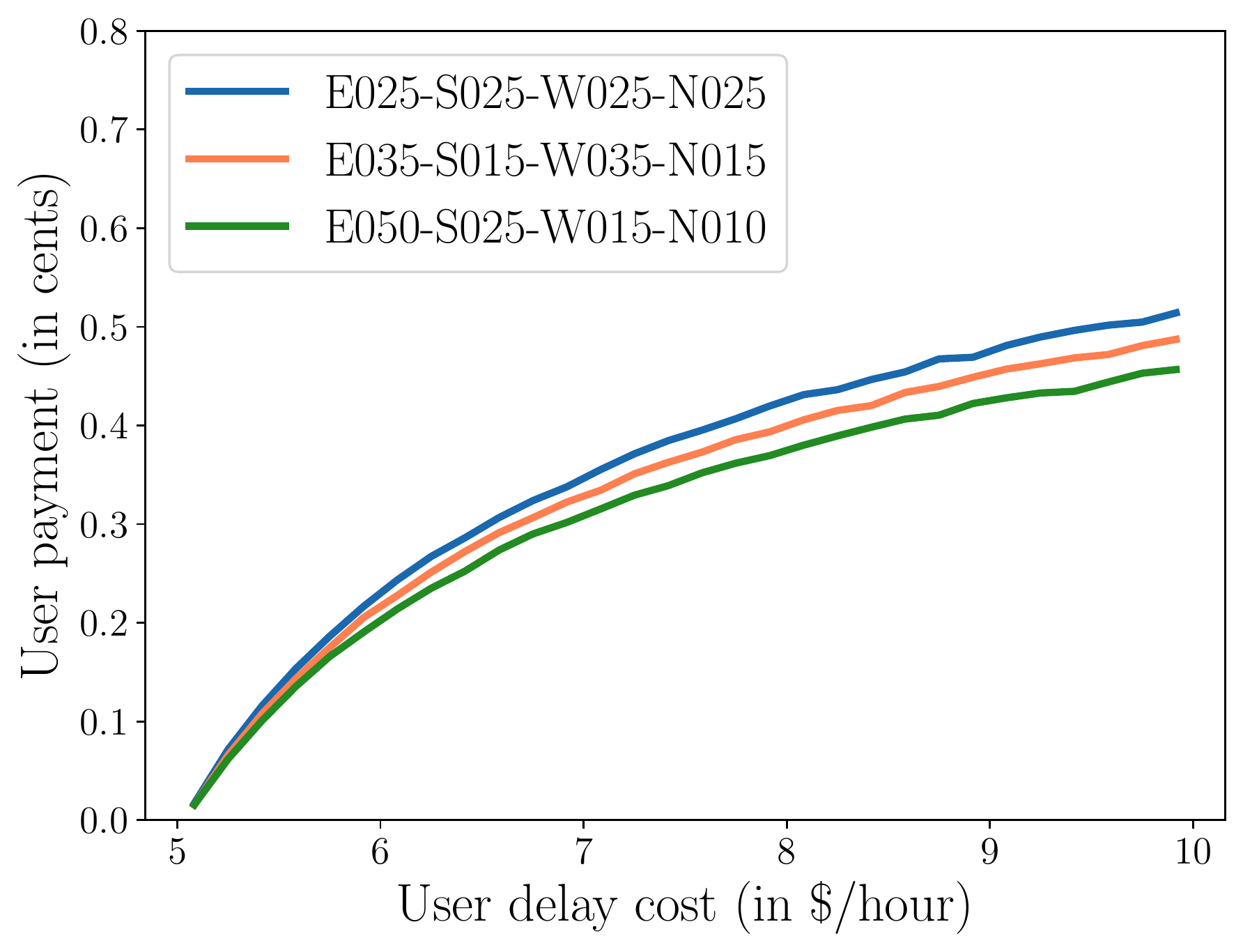}
			\caption{Average payment over all four lanes. \label{fig:payavglanes_queue}}
		\end{subfigure} &
		\begin{subfigure}{0.35\linewidth}
			\centering\includegraphics[width=1.0\linewidth]{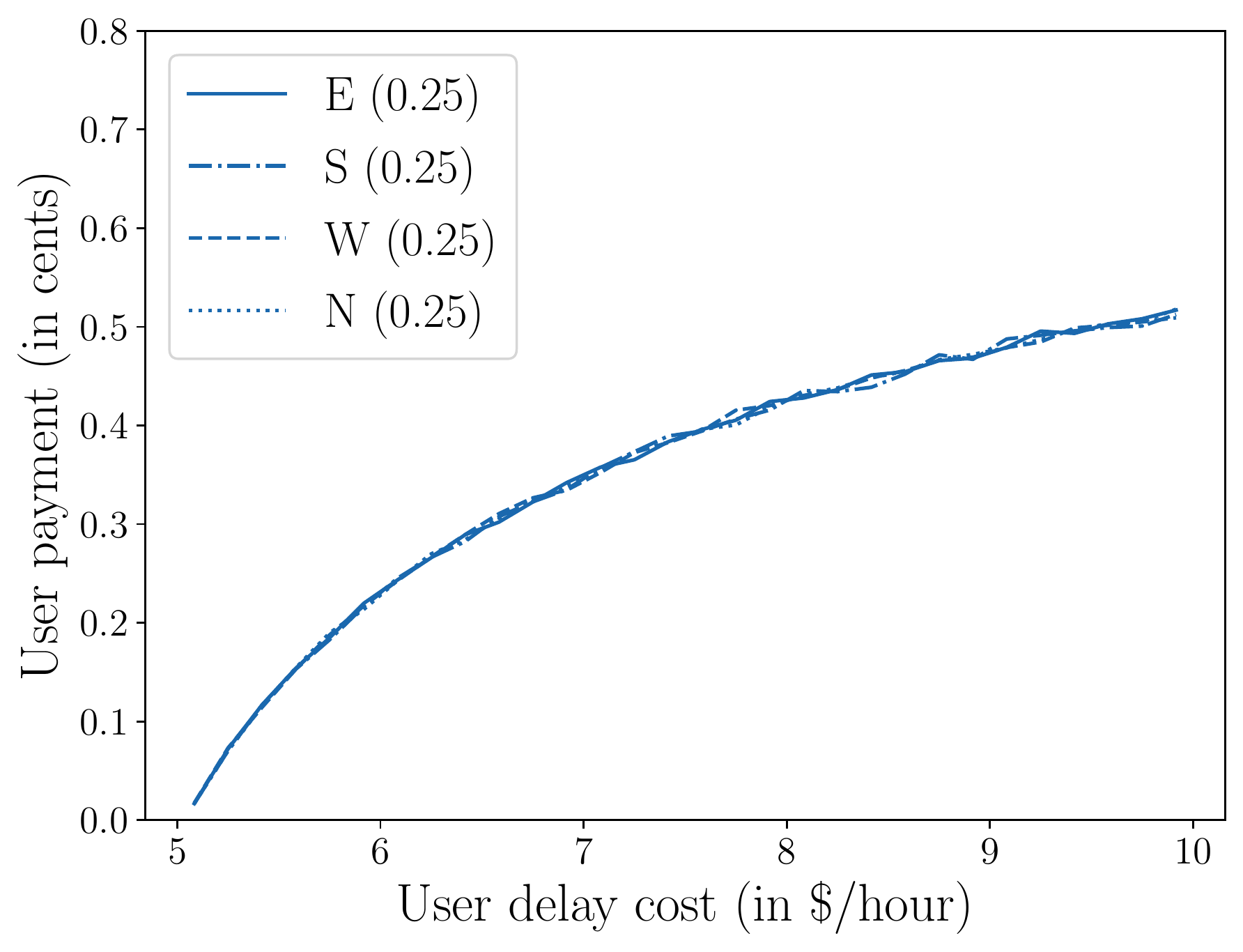}
			\caption{Lane-based payment for E025-S025-W025-N025.\label{fig:pay025_queue}}
		\end{subfigure} \\
		\begin{subfigure}{0.35\linewidth}
			\centering\includegraphics[width=1.0\linewidth]{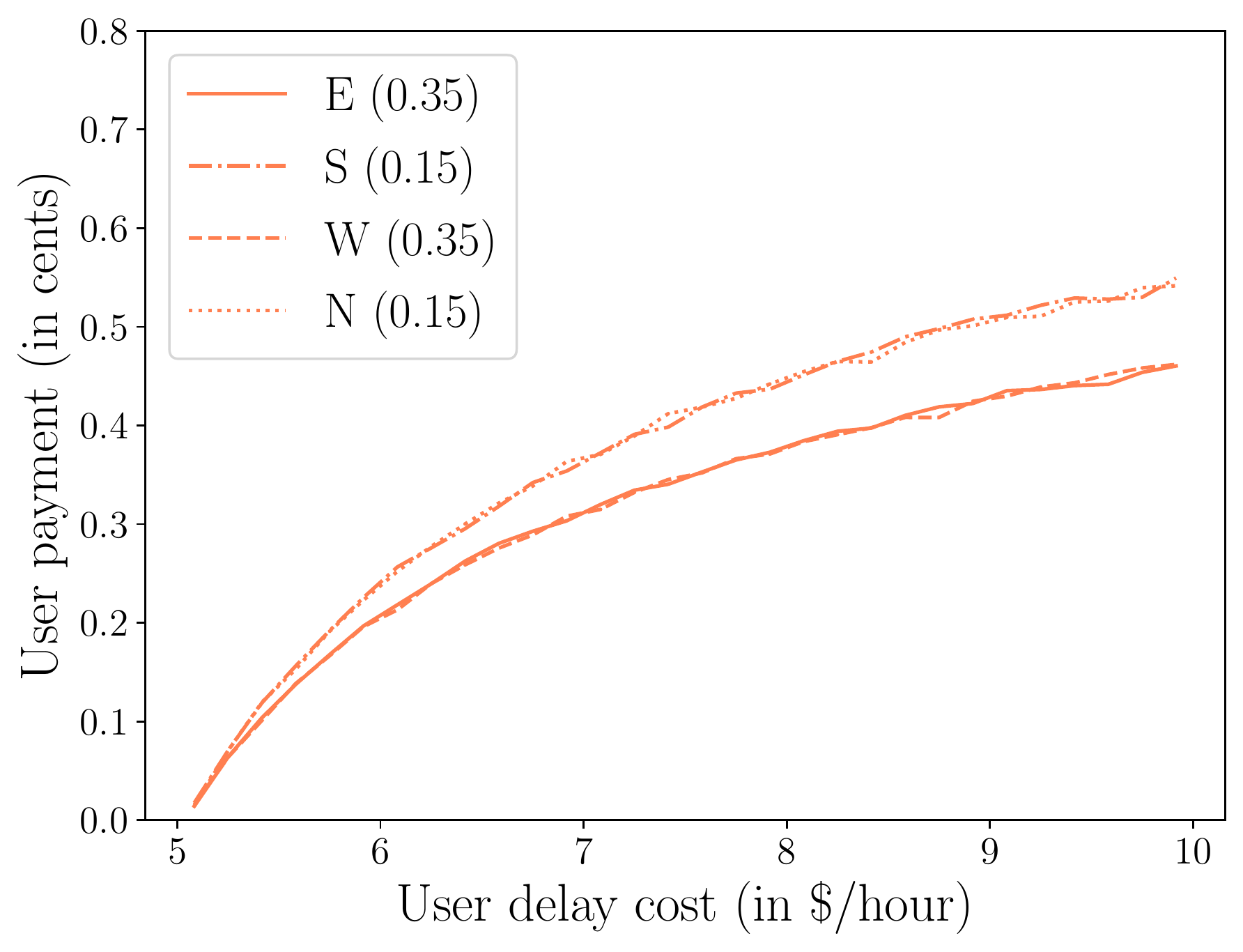}
			\caption{Lane-based payment for E035-S015-W035-N015.\label{fig:pay035_queue}}
		\end{subfigure} &
		\begin{subfigure}{0.35\linewidth}
			\centering\includegraphics[width=1.0\linewidth]{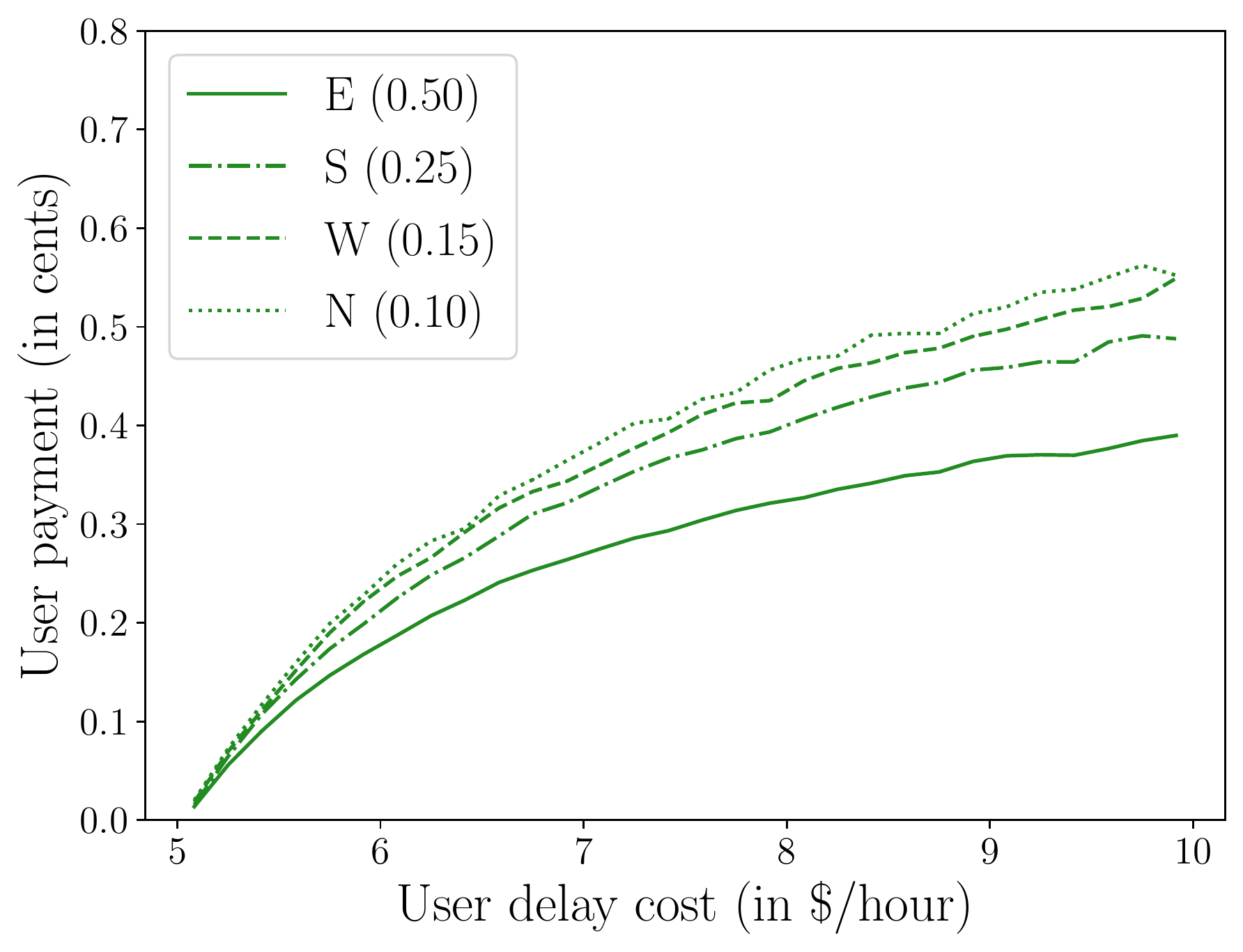}
			\caption{Lane-based payment for E050-S025-W015-N010.\label{fig:pay050_queue}}
		\end{subfigure} \\
	\end{tabular}
	\caption{Average user payments using the queue-based mechanism. Results of three simulations with varying lane-based arrival probabilities $p_j$, $j\in\{\text{E,S,W,N}\}$. Each simulation consists of 1 million users.\label{fig:paylane_queue}}
\end{figure}

\begin{figure}[!h]
	\centering
	\begin{tabular}{cc}
		\begin{subfigure}{0.35\linewidth}
			\centering\includegraphics[width=1.0\linewidth]{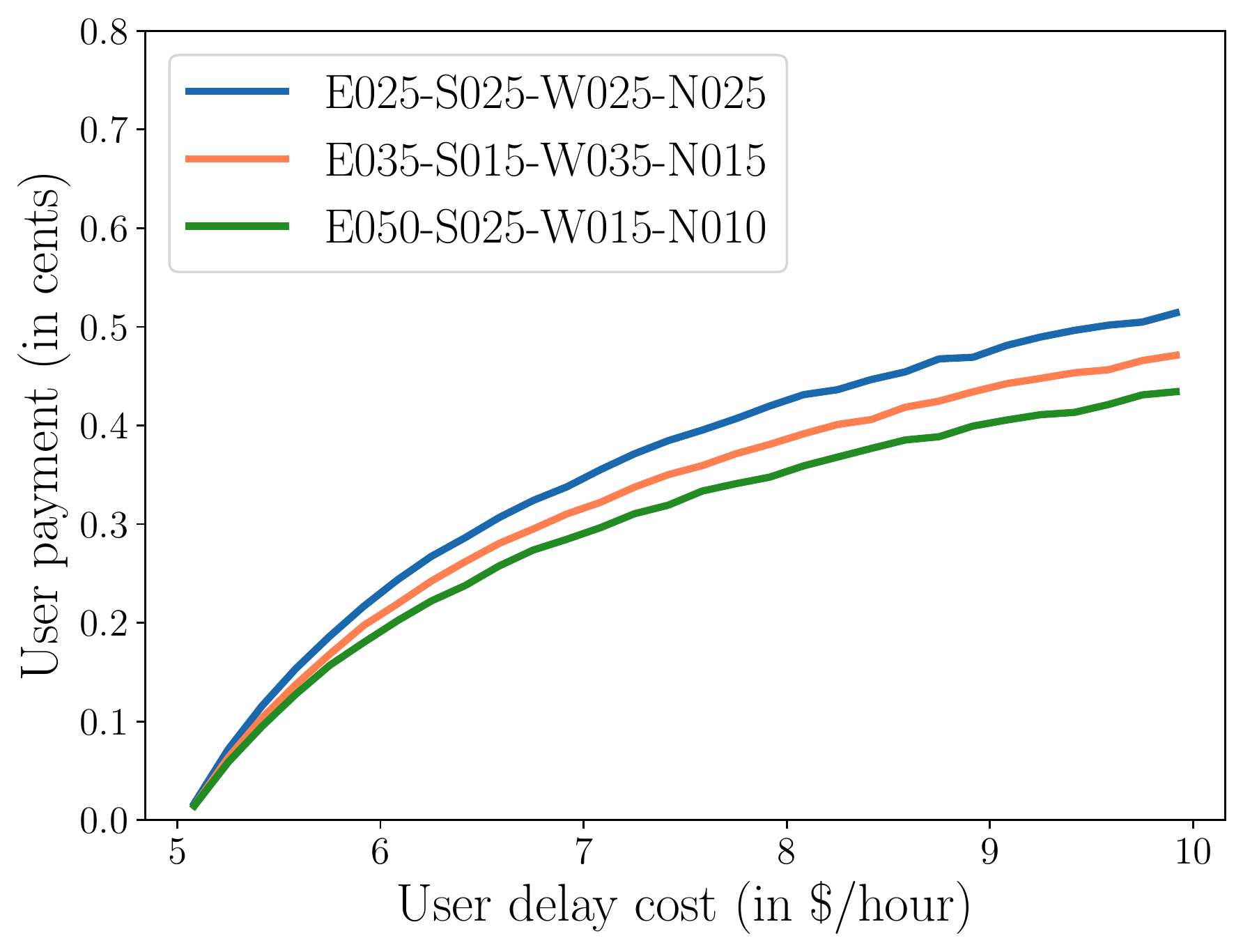}
			\caption{Average payment over all four lanes. \label{fig:payavglanes}}
		\end{subfigure} &
		\begin{subfigure}{0.35\linewidth}
			\centering\includegraphics[width=1.0\linewidth]{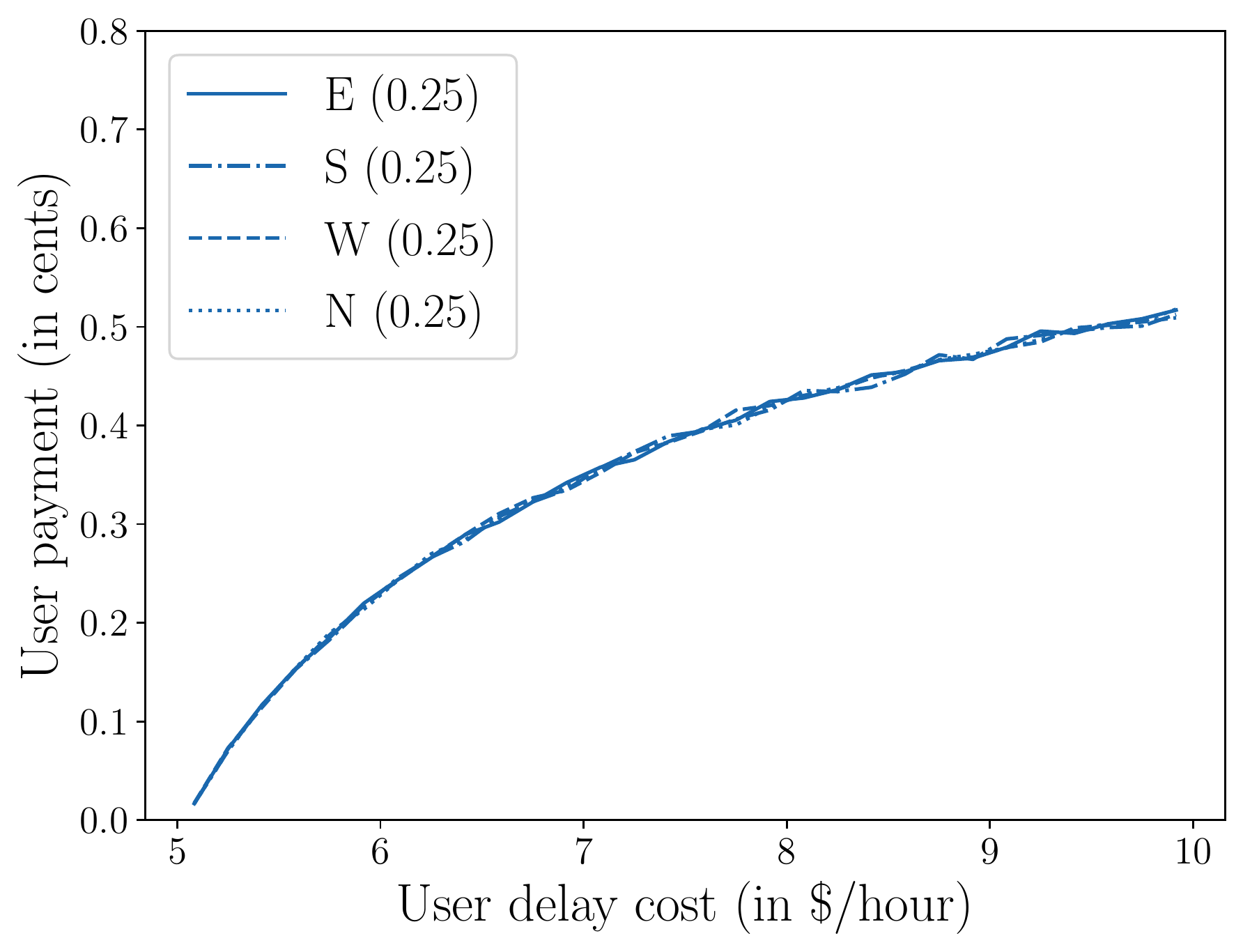}
			\caption{Lane-based payment for E025-S025-W025-N025.\label{fig:pay025}}
		\end{subfigure} \\
		\begin{subfigure}{0.35\linewidth}
			\centering\includegraphics[width=1.0\linewidth]{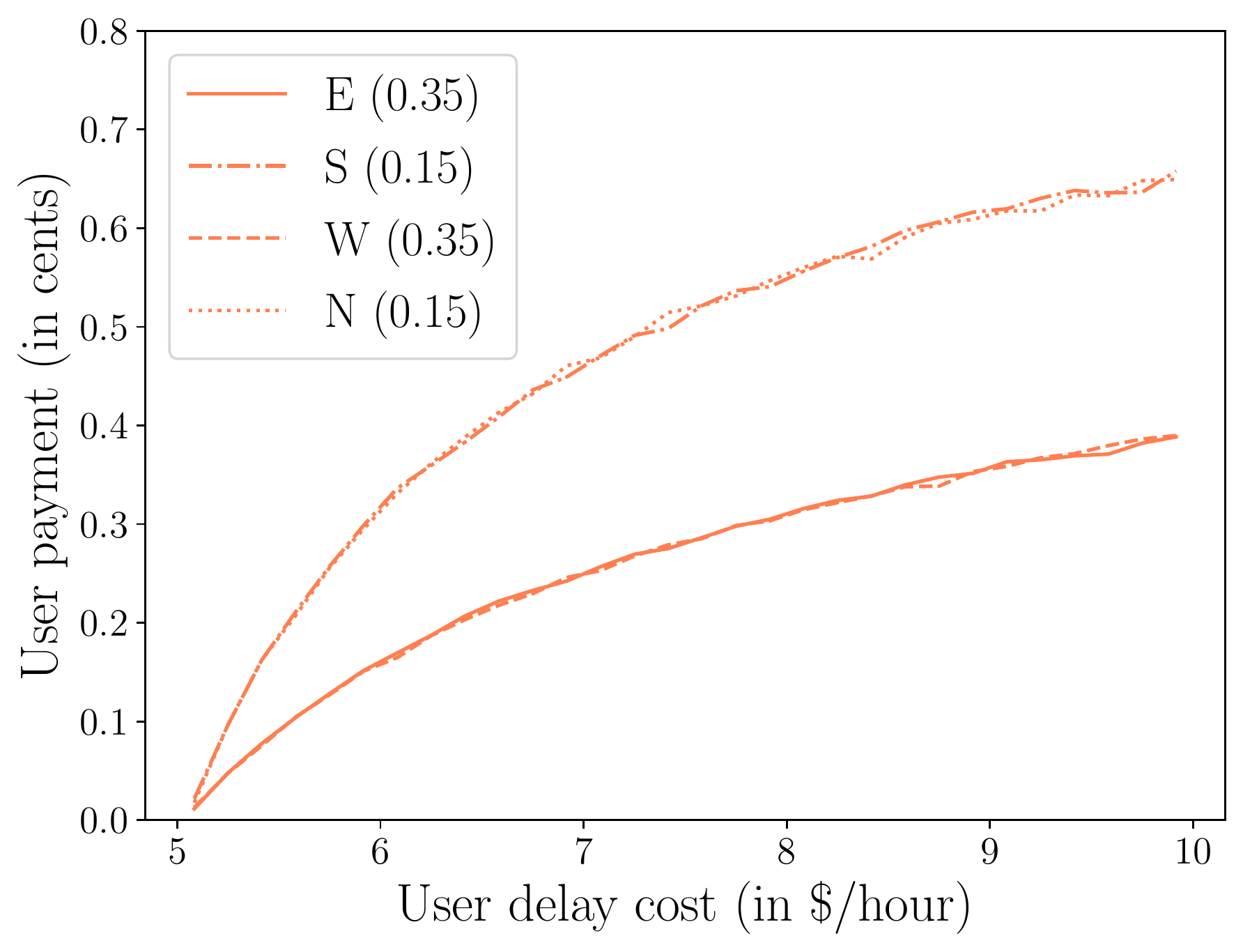}
			\caption{Lane-based payment for E035-S015-W035-N015.\label{fig:pay035}}
		\end{subfigure} &
		\begin{subfigure}{0.35\linewidth}
			\centering\includegraphics[width=1.0\linewidth]{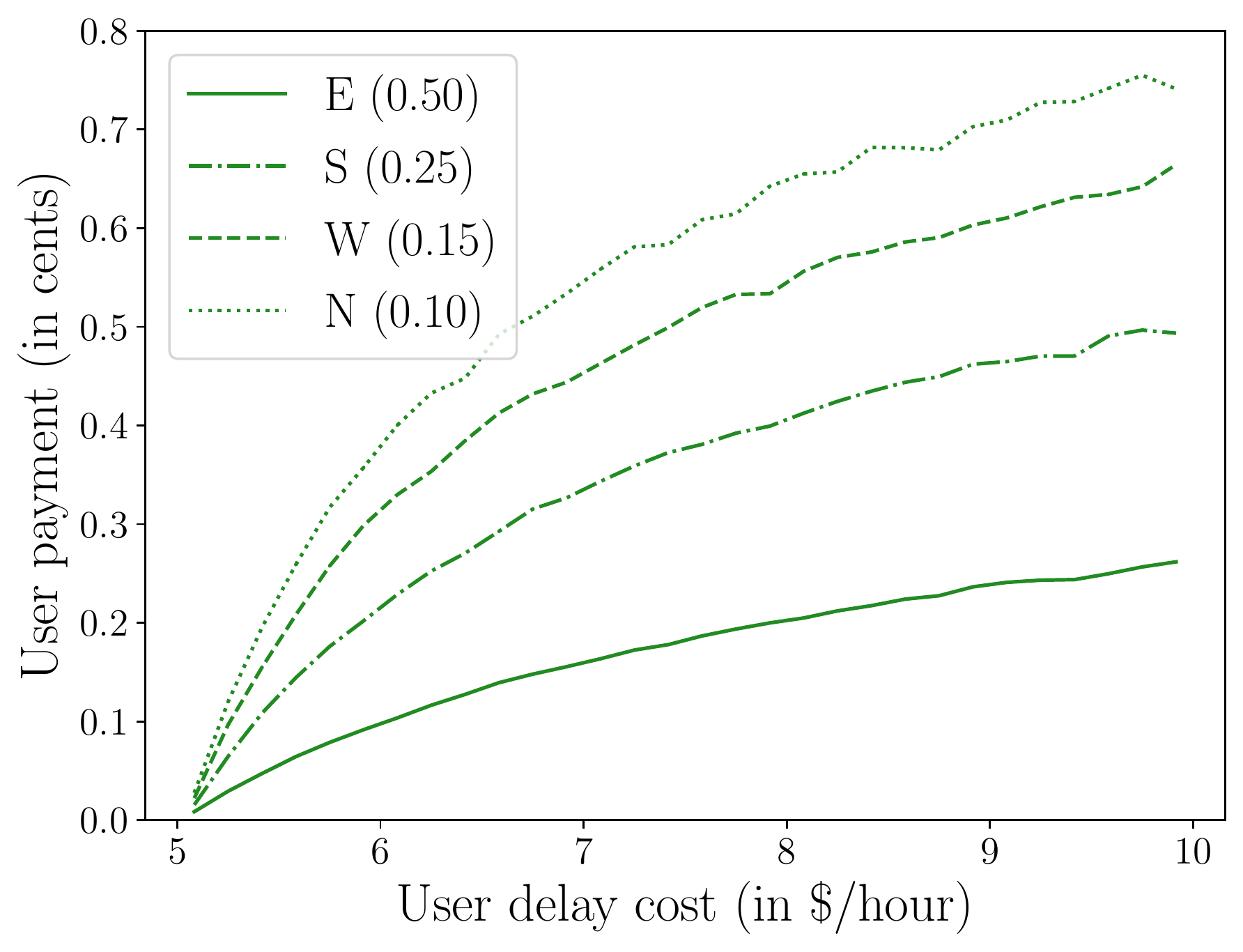}
			\caption{Lane-based payment for E050-S025-W015-N010.\label{fig:pay050}}
		\end{subfigure} \\
	\end{tabular}
	\caption{Average user payments using the lane-based mechanism. Results of three simulations with varying lane-based arrival probabilities $p_j$, $j\in\{\text{E,S,W,N}\}$. Each simulation consists of 1 million users.\label{fig:paylane}}
\end{figure}



We next further examine the behavior of the proposed mechanisms in the context of non-uniform lane arrival probabilities. We compare three configurations which correspond to an expected arrival rate of users of 1. The first configuration consists of uniform lane arrival probabilities of $p_j = 0.25$ for all lane $j \in \{\text{E,S,W,N}\}$. The second configuration consists of an arrival probability of $p_j = 0.35$ for East and West lanes, and $p_j = 0.15$ for South and North lanes. The third configuration consists of fully asymmetric lane arrival probabilities of $0.50$, $0.25$, $0.15$ and $0.10$ for lanes East, South, West and North, respectively. These three configurations are denoted Ewww-Sxxx-Wyyy-Nzzz where the three digits next to the lane direction represent the associated probability with two decimal values, e.g. E025 correspond to $p_E=0.25$. For each of the three configurations tested, we simulate the arrival of 1 million users at an intersection.

Figure \ref{fig:awtlane} depicts the average experienced user waiting time in the three lane arrival probabilities configurations. The top left sub-figure (\ref{fig:awtavglanes}) shows the average behavior over all four lanes of the intersection, whereas sub-figures (\ref{fig:awt025}-\ref{fig:awt050}) illustrate lane-based experienced waiting time for each of the three configurations tested. We find that the uniform configuration (E025-S025-W025-N025 -- Figure \ref{fig:awt025}) leads to higher average waiting times compared to the other two configurations tested. As expected, the uniform configuration yields a uniform waiting time all lanes. In contrast, non-uniform configurations lead to lane-specific waiting times. The outcome of the partially asymmetric configuration (E035-S015-W035-N015 -- Figure \ref{fig:awt035}) reveals that experienced waiting times for the lower arrival probability lanes (South and North) are marginally higher to that of the uniform configuration (E025-S025-W025-N025), whereas waiting times for the higher arrival probability lanes (East and West) are substantially lower. Results for the fully asymmetrical configuration (E050-S025-W015-N010 -- Figure \ref{fig:awt050}) show that waiting times are higher on low arrival probability lanes and decrease with the arrival probability.

Figures \ref{fig:paylane_static}, \ref{fig:paylane_queue} and \ref{fig:paylane} depict the variations of average user payments using the static, queue- and lane-based mechanisms, respectively. Similarly to the layout of Figure \ref{fig:awtlane}, the top-left sub-figure illustrates the average user payment over all lanes of the intersection, whereas the other three sub-figures illustrate lane-specific average user payments for each of the three configurations tested. In all cases, we find that user payments increase with users' delay costs and that the variance across lanes of the intersection increases as well with users' delay costs. The static mechanism (Figure \ref{fig:paylane_static}) produces near linear user payments w.r.t. users' delay costs and yields a low variance across lanes, for both asymmetric lane arrival probabilities configurations tested (Figures \ref{fig:pay035_static} and \ref{fig:pay050_static}). In turn, the queue- and lane-based mechanisms generate concave-shaped user payments w.r.t. users' delay costs. Further, for the asymmetric configurations, we observe that the queue-based mechanism (Figures \ref{fig:pay035_queue} and \ref{fig:pay050_queue}) yields lane-specific user payments that have a lower variance compared to the user payments obtained using the lane-based mechanism (Figures \ref{fig:pay035} and \ref{fig:pay050}). Specifically, for the configuration E035-S015-W035-N015, user payments for the highest user delay costs are comprised between 0.4~\textcent\ and 0.55~\textcent\ using the queue-based mechanism. Using the lane-based mechanism, these figures change to 0.35~\textcent\ and 0.65~\textcent. For the configuration E050-S025-W015-N010, user payments are lower on high arrival probability lanes and increase on less demanded lanes. This can be explained by observing that users on low arrival probability lanes have a higher chance of delaying more users since those are expected to arrive more frequently on higher arrival probability lanes.

\subsection{Computational runtime analysis}
\label{runtime}

\begin{figure}
	\centering
	\begin{subfigure}{0.45\linewidth}
		\centering\includegraphics[width=\columnwidth]{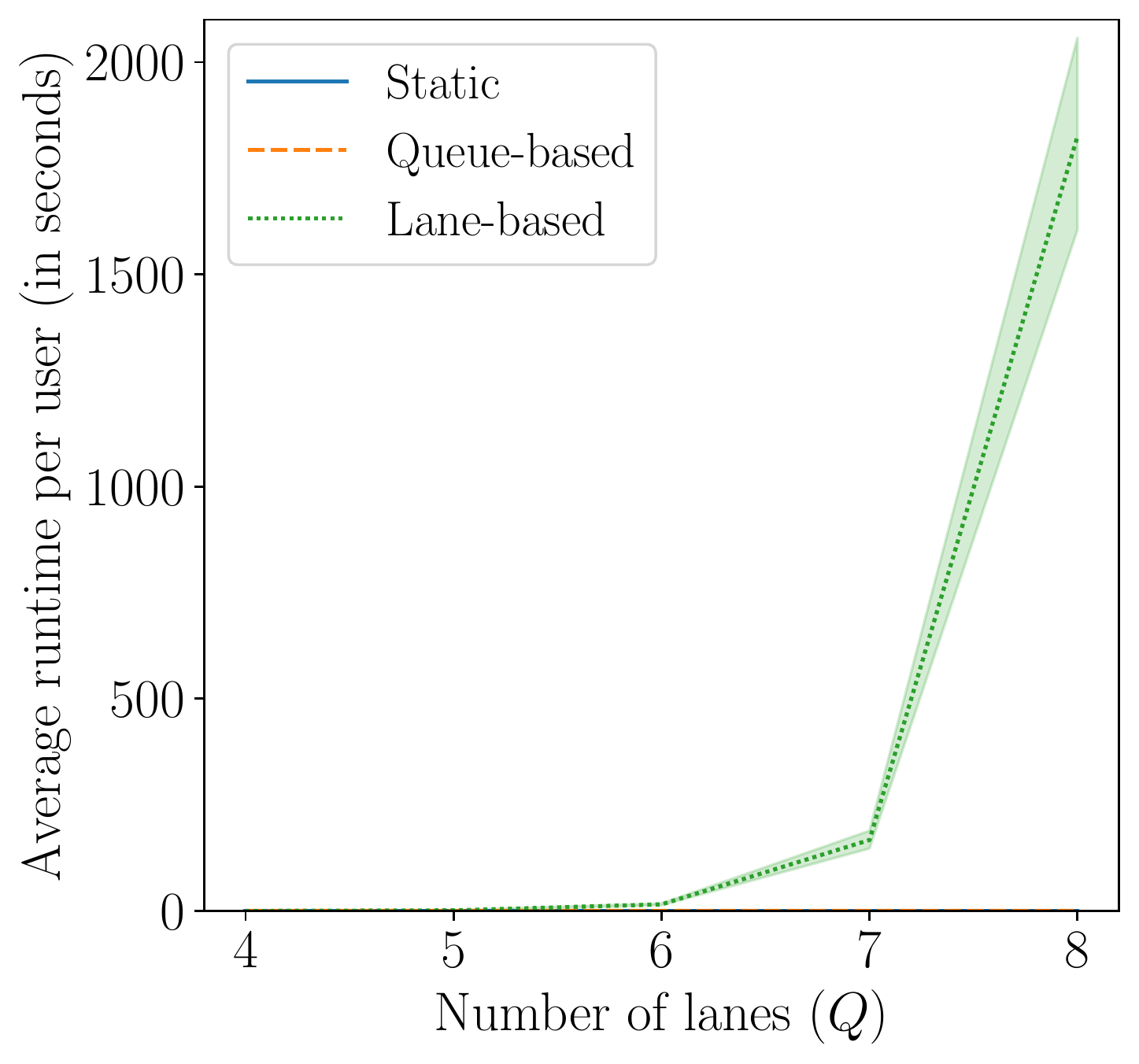}
		\caption{Mechanisms runtime.\label{fig:rt1}}
	\end{subfigure}
	\begin{subfigure}{0.45\linewidth}
		\centering\includegraphics[width=\columnwidth]{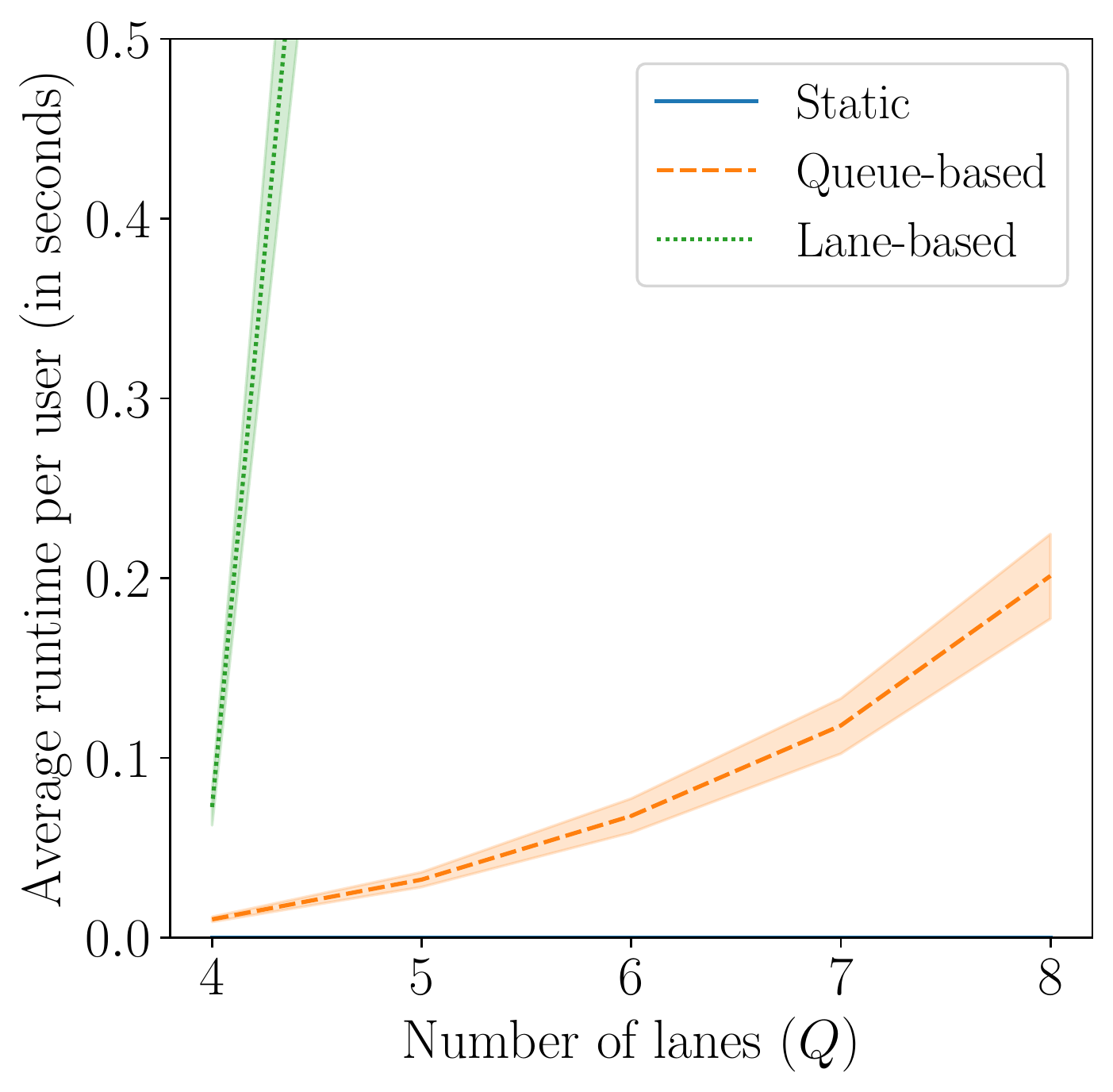}
		\caption{A closer look.\label{fig:rt2}}
	\end{subfigure}
	\caption{Average runtime per user for varying number of lanes ($Q$) in the intersection for each mechanism (static, queue-based and lane-based). The data for each number of lanes corresponds to a simulation with 100 users. Figure \ref{fig:rt2} is a closer look at Figure \ref{fig:rt1}. The error envelope represents the 95\% confidence interval of each data series.\label{fig:rt}}
\end{figure}

To analyses the performance of the proposed traffic auction mechanisms across varying intersection sizes, we conduct a sensitivity analysis by increasing the number of lanes $Q$ from 4 to 8 and report the average computational runtime over 100 users for all three mechanisms (static, queue-based and lane-based) per user in Figure \ref{fig:rt}. We also report the number of states for each intersection size in both the queue- ($\left\vert\Siq\right\vert$) and the lane-based ($\left\vert\Siz\right\vert$) in Table \ref{tab:states}. We find that the runtime of the static mechanism is of the order of 0.0001 s and near constant with regards to the intersection size. The average runtime per user using the queue-based mechanism is of the order of magnitude of 0.01 s for $Q=4$ and increases up to nearly 0.1 a for $Q=8$. In contrast, the average runtime per user required by the lane-based mechanism increases exponentially with the number of lanes. The average runtime per user with $Q=4$ is slightly less 0.1 s, and increases to over 10 s for $Q=6$ lanes. For an 8-lane intersection, the average runtime per user exceeds 1000 s, which can be explained by the substantially larger number of states in the lane-based model, $\left\vert\Siz\right\vert = 2187$, compared to the queue-based model, $\left\vert\Siq\right\vert = 36$, as reported in Table \ref{tab:states}.

\begin{table}
\begin{tabular}{lll}
\toprule
$Q$ & $\left\vert\Siq\right\vert$ & $\left\vert\Siz\right\vert$ \\
\midrule
4 & 10 & 27 \\
5 & 15 & 81 \\	
6 & 21 & 243 \\		
7 & 28 & 729 \\			
8 & 36 & 2187 \\					
\bottomrule		
\end{tabular}
\caption{Number of states in the queue- ($|\Siq|$) and lane-based ($|\Siz|$) models, for varying number of lanes $Q$ in the intersection. \label{tab:states}}
\end{table}

\section{Conclusions}
\label{con}

We next summarize the findings of this study before discussing its limitations and outlining potential extensions and future research directions.

\subsection{Summary of findings}

In this paper, we have presented novel online mechanisms for traffic intersection auctions. We assume that users can declare their delay cost privately to the intersection manager and focus on determining incentive-compatible payments for users at the front of their lane. We also assume that the intersection manager has knowledge of the distribution of users' delay cost. The proposed mechanisms are designed to minimize users' generalized cost which is defined as a linear combination of expected waiting time and user payment. We introduced two Markov chain models to determine users' expected waiting time, and presented a payment mechanism that can be implemented with both models. We showed that the proposed online traffic intersection mechanisms are incentive-compatible in the dynamic sense and thus maximize social welfare. We conducted numerical experiments on a four-lane traffic intersection to explore the behavior of the proposed online mechanisms and compared their performance to that of a static incentive-compatible mechanism. Our findings highlight that under the auction configuration tested, i.e. uniform user delay costs along with a four-lane traffic intersection, static auction mechanisms may not be incentive-compatible in dynamic sense. That is, static incentive-compatible mechanisms (which do not account for future arrivals) may fail to ensure truthful user behavior by incentivizing users to misreport their delay cost in the long run. We also provide empirical evidence that the proposed online mechanisms are computationally efficient and could be used to manage traffic operations in real-time. We quantified the trade-offs between the queue- and lane-based mechanisms and verified that the lane-based mechanism can provide lane-specific incentive-compatible payments. Further, we examined the influence of the number of lanes onto the computational scalability of the proposed online mechanisms. Our experiments show that while the runtime of both dynamic mechanisms grow exponentially with the number of lanes, the average per-user runtime of the queue-based mechanism remains competitive even for large, i.e. eight-lane traffic intersections; whereas the lane-based mechanism requires considerable computational resources beyond seven lanes. This highlights the trade-off between computational scalability and model accuracy and suggests that if lane arrival probabilities are uniform or near-uniform, the queue-based provides a scalable solution for real-time pricing. On the other hand, the lane-based model is better suited for traffic intersections with non-uniform lane arrival probabilities. 

Compared to the existing literature on traffic intersection auctions, to the best of our knowledge, this paper is the first to propose incentive-compatible mechanisms that account for future user arrivals within the auction process. This improves on prior works that are restricted to static settings in which auction rounds are run sequentially and during which new arrivals may further delay auction losers, potentially violating the incentive-compatibility property in the long run.

\subsection{Limitations and future research directions}

We next discuss the assumptions and mechanism design choices made in this research. 

The proposed online mechanisms focus on determining incentive-compatible payments for users at the front of their lane queues. Since all users queueing on a lane of a traffic intersection are eventually going to reach the front of their lanes, all users will eventually join the pricing queue and be required to declare their delay cost so as to obtain their corresponding payment. However, the proposed mechanisms do not provide the option for ``blocked'' users, i.e. users queueing behind another user on their lane, to ``vote'' for the user at the front of their lane queues. Although this is a limitation of the proposed online mechanisms, since blocked users are assumed to be unaware of other users' delay cost, it appears challenging to ensure that they could reap benefits from bidding for their lane leader. 

The geometry of traffic intersections is only implicitly accounted for within the proposed online mechanisms. Specifically, users' service times are captured within the one-step-costs of the proposed Markov chain models used to determine the expected waiting time of users. Yet, in our numerical experiments, we have assumed that all service times are deterministic and uniform. The uniformness assumption could be relaxed by substituting the unit one-step costs with movement-specific service times. However, relaxing this assumption would also require the introduction of a detailed model of users' travel time within the intersection which is function of other users' route choice and service time. This is expected to require an extension of the state space in the proposed lane-based Markov chain model to a movement-based model. Further, the simulation of realistic traffic movements within the intersection is expected to require traffic control model that provides the flexibility of servicing in signal-free context. Such models have been widely explored in the connected and autonomous vehicles' literature and we leave the extension of the proposed online mechanisms to realistic traffic models for future works. The deterministic service time assumption could be relaxed at the expense of introducing more complex modeling of users' trajectory interactions within the intersection. Alternatively, continuous-time Markov chains for which the transition time is random or discrete-time Markov chains where the arrival probability varies depending on the elapsed time due to service might handle non-uniform service times.

The proposed research may also be extended by considering batches or coalitions of users. This may lead to improved social benefits if, for instance, platoons of vehicles are priced to traverse the intersection in a effective manner. Such market-driven approaches could be combined with recent efforts on platoon coordination at traffic intersections \citep{jiang2006platoon,lioris2017platoons,monteil2018mathcal,niroumand2020joint}. The proposed online mechanisms could be revisited under the light of  fairness. As observed by \citet{schepperle2007agent} ``starvation'' effects may occur if intersection access is blocked by low delay cost users or dominated by high delay cost users on specific lanes. Future research is thus needed to advance the development of market-driven mechanisms that are able to combine dynamic incentive-compatibility with equity factors. Finally, methods on load balancing in parallel queueing systems as proposed by \citet{down2006dynamic} could be used to develop lane assignment policies to assign users to small queue-length lanes so as to improve the social welfare of the system. 

\bibliography{DR}

\end{document}